\documentclass[twocolumn,showpacs,preprintnumbers,aps,pra,amsmath,amssymb,floatfix,superscriptaddress]{revtex4-1}

\usepackage{color}
\usepackage{graphicx}      
\usepackage{subfigure}
\usepackage{dcolumn}
\usepackage{textcomp}
\usepackage{longtable}     
\usepackage{url}           
\usepackage{bm}            
\usepackage{appendix}

\usepackage{braket}

\usepackage{hyperref}		
\hypersetup{colorlinks=true,breaklinks=true,urlcolor=blue,citecolor=blue,linkcolor=blue,pdfstartview=FitH,pdfpagemode=UseNone} 

\begin{document}

\title{High Resolution Molecular Spectroscopy for Producing Ultracold Absolute Ground-State $^{23}$Na$^{87}$Rb Molecules}

\author{Mingyang Guo}
\email{myguo@phy.cuhk.edu.hk}
\affiliation{Department of Physics, The Chinese University of Hong Kong, Hong Kong, China}
\author{Romain Vexiau}
\affiliation{Laboratoire Aim\'e Cotton, CNRS, Universit\'e Paris-Sud, ENS Paris-Saclay, Universit\'e Paris-Saclay, 91405 Orsay Cedex, France}
\author{Bing Zhu} \thanks{Present address: Physikalisches Institut, Universit\"{a}t Heidelberg, INF 226, 69120 Heidelberg, Germany}
\affiliation{Department of Physics, The Chinese University of Hong Kong, Hong Kong, China}
\author{Bo Lu} \thanks{Present address: School of Physics and Astronomy, Sun Yat-Sen University, Zhuhai 519082, China}
\affiliation{Department of Physics, The Chinese University of Hong Kong, Hong Kong, China}
\author{Nadia Bouloufa-Maafa}
\affiliation{Laboratoire Aim\'e Cotton, CNRS, Universit\'e Paris-Sud, ENS Cachan, Universit\'e Paris-Saclay, 91405 Orsay Cedex, France}
\author{Olivier Dulieu}
\email{olivier.dulieu@u-psud.fr}
\affiliation{Laboratoire Aim\'e Cotton, CNRS, Universit\'e Paris-Sud, ENS Cachan, Universit\'e Paris-Saclay, 91405 Orsay Cedex, France}
\author{Dajun Wang}
\email{djwang@cuhk.edu.hk}
\affiliation{Department of Physics, The Chinese University of Hong Kong, Hong Kong, China}
\affiliation{The Chinese University of Hong Kong Shenzhen Research Institute, Shenzhen, China} 

\date{\today}

\begin{abstract}

We report a detailed molecular spectroscopy study on the lowest excited electronic states of $^{23}\rm{Na}^{87}\rm{Rb}$ for producing ultracold $^{23}\rm{Na}^{87}\rm{Rb}$ molecules in the electronic, rovibrational and hyperfine ground state. Starting from weakly-bound Feshbach molecules, a series of vibrational levels of the $A^{1}\Sigma^{+}-b^{3}\Pi$ coupled excited states were investigated. After resolving, modeling and interpreting the hyperfine structure of several lines, we successfully identified a long-lived level resulting from the accidental hyperfine coupling between the $0^+$ and $0^-$ components of the $b^3\Pi$ state, satisfying all the requirements for the population transfer toward the lowest rovibrational level of the X$^1\Sigma^+$ state. Using two-photon spectroscopy, its binding energy was measured to be 4977.308(3)~cm$^{-1}$, the most precise value to date. We calibrated all the transition strengths carefully and also demonstrated Raman transfer of Feshbach molecules to the absolute ground state. 
	
\end{abstract}

\maketitle

\section{Introduction}
\label{sec:intro}

In a recent paper, the authors reported the successful formation of a gaseous cloud of ultracold $^{23}$Na$^{87}$Rb ground-state molecules in an optical trap~\cite{guo2016creation}. This achievement represents a new step in the quest for one of the current major challenge of atomic, molecular and optical physics: obtaining a dense gas of ultracold polar particles suitable for the observation of strong anisotropy of their mutual interaction. By \textit{polar particle} we mean a paramagnetic atom (like Cr, Er, Dy,...) or a molecule possessing a magnetic dipole moment or a permanent electric dipole moment in its own frame, or both. In this context, the $^{23}$Na$^{87}$Rb molecule in its $X^1\Sigma^+$ electronic ground state has no magnetic moment, and is characterized by one of the strongest electric permanent dipole moment (3.2~D) \cite{guo2016creation} among the series of ten bialkali heteronuclear species that can be created out of the association of two different alkali-metal atoms among Li, Na, K, Rb, Cs \cite{aymar2007calculations}. Moreover the various atomic isotopes of the alkali-metal atoms allows for creating ``on demand'' either bosonic or fermionic molecules, which determines the nature of their quantum degeneracy, namely the formation of a Bose-Einstein condensate or of a quantum degenerate Fermi gas. Last but not least, only five species out this series of ten (NaK, NaRb, NaCs, KCs, RbCs) are stable when two identical molecules in their absolute ground state collide together. Thus, such ultracold polar molecules are promising candidates to investigate ultracold dipolar many-body physics and novel quantum phases \cite{baranov2008theoretical,lahaye2009physics,yi2007novel} resulting from their anisotropic interactions when they are exposed to an appropriate external electric field.

So far, the most successful method to create ultracold bialkali polar molecules proceeds through two main steps.  First a very weakly-bound ground state molecule is created by associating ultracold atoms across a Feshbach resonance \cite{kohler2006production,chin2010feshbach}. A stimulated Raman adiabatic passage (STIRAP) \cite{bergmann1998coherent} is then applied to transfer the weakly-bound Fesbhach molecules to their singlet ground state, in which the molecules possess large permanent electric dipole moments. This scheme has been successfully applied for creating  fermionic molecules of KRb \cite{ni2008high} and NaK \cite{park2015ultracold}, as well as bosonic samples of RbCs molecules \cite{takekoshi2014ultracold, molony2014creation}. The same method has been used in Ref.~\cite{guo2016creation} to create ultracold bosonic $^{23}$Na$^{87}$Rb molecules. 

\begin{figure}[!t]
	\centering
	\includegraphics[width=0.4\textwidth]{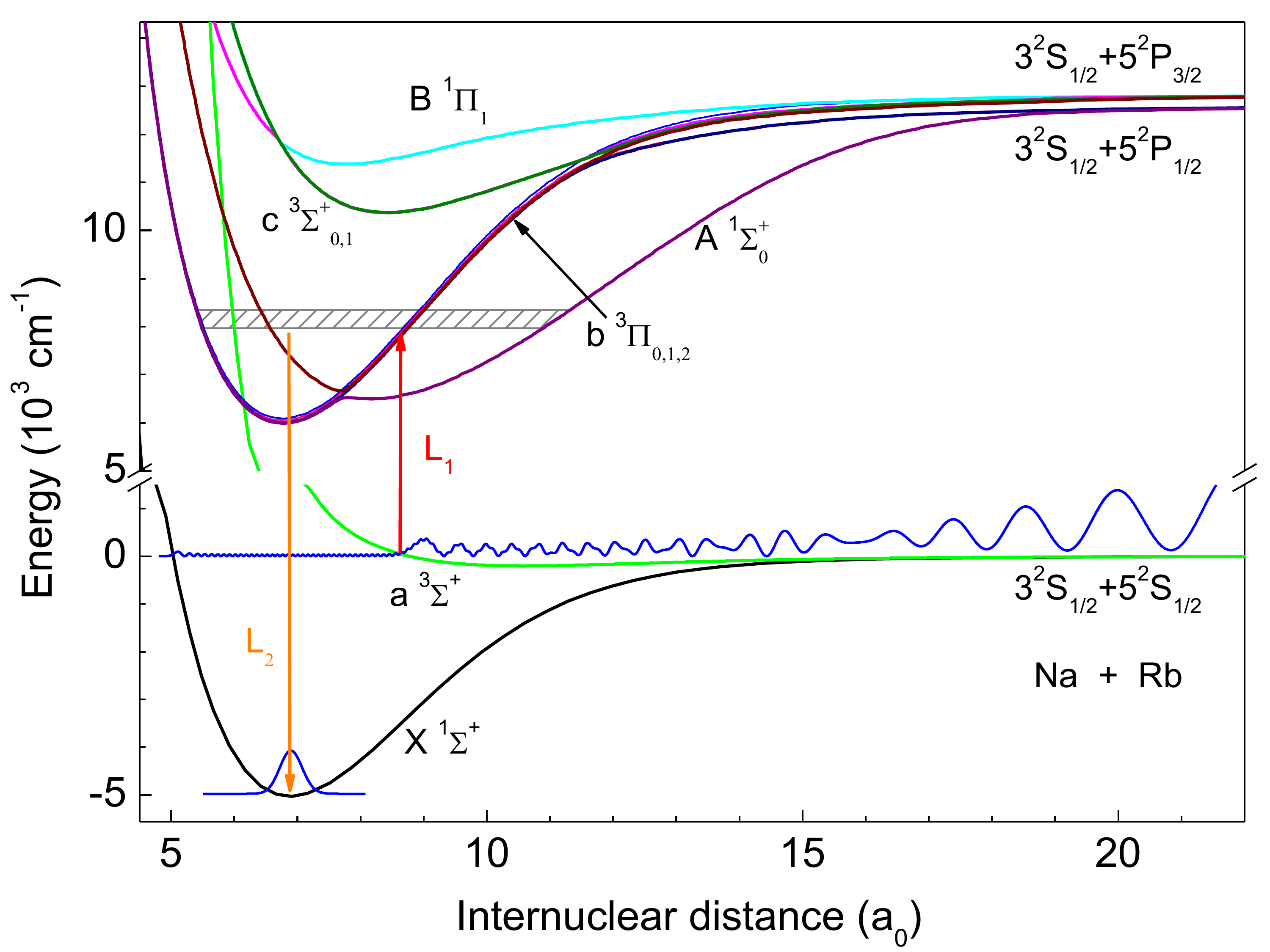}
	\caption {The two-photon path used in this work to create ground-state $^{23}$Na$^{87}$Rb molecules, and the relevant $^{23}$Na$^{87}$Rb potential energy curves including spin-orbit coupling involved in the proposed three-level system. The $\ket{1}$ state is the Feshbach state (for which squared radial probability density is drawn) with a dominant $a^3\Sigma^+$ character, and is coupled through the pump laser $L_1$ to the $\ket{2}$ level belonging to the strongly mixed $A^1\Sigma^+-b^3\Pi$ system. The dump laser $L_2$ couples $\ket{2}$ to the absolute NaRb ground state level $\ket{3}$. The dashed area represents the energy range explored in the experiment.}
	\label{fig1}
\end{figure}

In the present paper, we report on the details of the steps which lead us to this successful result, which are illustrated in Fig.~\ref{fig1} for the present $^{23}$Na$^{87}$Rb case. A typical STIRAP scheme involves three levels $\ket{1}$ (the weakly-bound level of the Feshbach molecule), $\ket{2}$ (a properly selected bound level of the excited state manifold of the molecule), and $\ket{3}$ (the absolute ground state level of $^{23}$Na$^{87}$Rb), coupled by two coherent laser pulses $L_1$ (\textit{pump} laser) and $L_2$ (\textit{dump} laser). With properly designed shape and duration of the pulses, a  model involving quantum states dressed by the electromagnetic field reveals a so-called ``dark state'', which is a superposition of states $\ket{1}$ and $\ket{3}$. Decoupled from the excited state, the dark state is freed from the fast spontaneous decay of the intermediate state $\ket{2}$ while allowing the complete transfer of the quantum state population from state $\ket{1}$ to state $\ket{3}$ \cite{bergmann1998coherent}. 

To achieve high transfer efficiency, the intermediate state $\ket{2}$ must satisfy several requirements. Firstly, state $\ket{2}$ should strongly couple to both state $\ket{1}$ and state $\ket{3}$. As the initial state $\ket{1}$ is ``triplet-dominated'' \cite{wang2015formation}, while state $\ket{3}$ is purely singlet, the intermediate states should have strong singlet-triplet mixing to achieve coupling with both $\ket{1}$ and $\ket{3}$ states. Furthermore, the Franck-Condon factors, representing the vibrational wavefunction overlap of state $\ket{2}$ with states $\ket{1}$ and $\ket{3}$, should be large enough to ensure strong couplings. On the other hand, the hyperfine structure of state $\ket{2}$ must be fully resolvable, \textit{i.e.} the splitting between hyperfine components must be larger than their natural linewidth, otherwise STIRAP can be spoiled by destructive interferences between different paths.

In heteronuclear alkali dimers, triplet-singlet mixing in excited states can typically be found in the $A^1\Sigma^+-b^3\Pi$ (or $A-b$ in short) and the $B^1\Pi-b^3\Pi-c^3\Sigma^+$ (or $B-b-c$ in short) systems (Fig.~\ref{fig1}). Here, we focus on the former system since it has been studied in detail with Fourier transform spectroscopy~\cite{docenko2007deperturbation}. However, we want to emphasize that high resolution spectroscopy down to the hyperfine structure, which is not available in Ref.~\cite{docenko2007deperturbation}, is necessary for the present purpose.  

Two types of mixed $A^1\Sigma^+-b^3\Pi$ energy levels are first investigated in this work, the so-called $A-b_{0^+}$ and the $A-b_1$ ones, where the 0 and 1 indexes refer to two distinct components of the $b$ manifold labeled with the value of the projection $\Omega=0, 1$ of the total electronic angular momentum (orbital + spin) on the molecular axis. The relevant coupling matrices were displayed in Ref. \cite{borsalino2016}. The hyperfine structure of the $A-b_{0^+}$ levels is found too weak for being properly resolved. The STIRAP transfer via the $A-b_1$ levels has been successfully implemented in RbCs \cite{debatin2011molecular,takekoshi2014ultracold}, but happens unsuccessful in the present work. Finally, a broad manifold of well-separated lines produced by the accidental mixing between the components $0^+$ and $0^-$ of the $b$ state (the plus and minus sign being related to the parity of the total electronic wavefunction with respect to the planar symmetry containing the molecular axis) is detected in the  explored spectroscopic region allows us to implement a novel scheme for an efficient STIRAP scheme.

This paper is organized as follows: In Sec.~\ref{section2}, we describe our experimental procedure to perform our spectroscopic measurements in the presence of a magnetic field. A coarse spectrum covering 15 vibrational levels is obtained, and modeled using up-to-date molecular structure data.  The hyperfine structures of three selected lines exhibiting characteristic substructures are presented in Sec.~\ref{section3}. An effective Hamiltonian model is set up to interpret these structures, revealing the specific sublevel which is appropriate for STIRAP. Using this intermediate level, we study in Sec.~\ref{section4} the structure of the rovibrational ground state with two-photon spectroscopy. More details on the derivation of matrix elements of the effective Hamiltonian, and on the obtained characteristics (transition strengths, radiative lifetimes, ...) of the hyperfine manifold of the intermediate level, are provided in the Appendix.

\section{Spectroscopy of the $^{23}$N\lowercase{a}$^{87}$R\lowercase{b} excited states}
\label{section2}

\subsection{Experimental procedure}

Our experiment starts from a sample of weakly-bound NaRb Feshbach molecules created via magneto-association involving a Feshbach resonance between the $\ket{F=1,m_F=1}$ hyperfine Zeeman states of Na and $^{87}$Rb atoms at 347.7~G~\cite{wang2013observation,wang2015formation}. Starting from a mixture of $1.5\times10^5$ Na atoms and $1.5\times10^5$ Rb atoms at a temperature of about 350~nK, about 10$^4$ pure Feshbach molecules are obtained after ramping the magnetic field across the Feshbach resonance and removing the residual atoms~\cite{wang2015formation}. At the final magnetic field of 335.6~G, the trap lifetime of these Feshbach molecules is longer than 20~ms.    

Both $L_1$ and $L_2$ lasers are provided by external cavity diode lasers stabilized to a dual-wavelength coated ultrastable optical cavity with measured finesses over 25000 at both wavelengths. The short-term laser linewidths are estimated to be less than 5~kHz and the long-term drift is less than 200~kHz per day. For fine frequency control, we inserted acoustic-optical modulators (AOMs) in multipass configurations between each laser and the ultrastable cavity. The radio frequency signals driving these AOMs are generated by signal generators with Hertz-level frequency resolutions. By changing the output frequencies of the signal generators, we can control the relative laser frequencies with a resolution of several kHz for a total scan range of up to 2~GHz. On the other hand, the absolute laser frequencies are measured by a calibrated wavelength meter with an accuracy of about 60~MHz. 

The two laser beams are delivered at the vicinity of the vacuum chamber with optical fibers. They are combined by a dichroic mirror and focused onto the molecular sample along a direction perpendicular to the vertical quantization axis defined by the magnetic field. Thus, each beam can selectively drive $\pi$ transitions with vertical polarization or $\sigma^{\pm}$ transitions with horizontal polarization.

Once the Feshbach molecule sample is prepared, we pulse on $L_1$ and measure the induced loss of Feshbach molecules vs. the $L_1$ frequency. At each $L_1$ frequency, a new molecular sample has to be prepared which takes about 45 seconds. Due to this low duty cycle, we first perform a coarse frequency scan with long laser pulse at the highest power. Once an excitation resonance is located, we scan the $L_1$ frequency with finer steps using shorter pulse duration and lower power to reveal the detailed structures. For the detection, we dissociate the Feshbach molecules into Na and Rb atoms and probe them with standard absorption image method. For this work, all the data are taken by imaging Rb atoms because of the better signal-to-noise ratio.

\subsection{Modeling the pump and dump transitions}

The efficiency of the transfer relies on the identification of a pair of pump and dump electric-dipole-allowed transitions with comparable Rabi frequencies, and thus on the detailed knowledge of the structure and the spectroscopy of the molecule of interest. The spectroscopy of the $X^1\Sigma^+$ and $a^3\Sigma^+$ states dissociating to Na(3$s$) + Rb(5$s$) has been carried on experimentally allowing precise determination of their potential energy curves (PEC) \cite{wang2013observation}.  The $A$ and $b$ PEC and their spin-orbit (SO) coupling are available from the extensive spectroscopic measurements and deperturbation analysis of Ref.~\cite{docenko2007deperturbation}.

\begin{figure}[!t]
	\centering
	\includegraphics[width=0.4\textwidth]{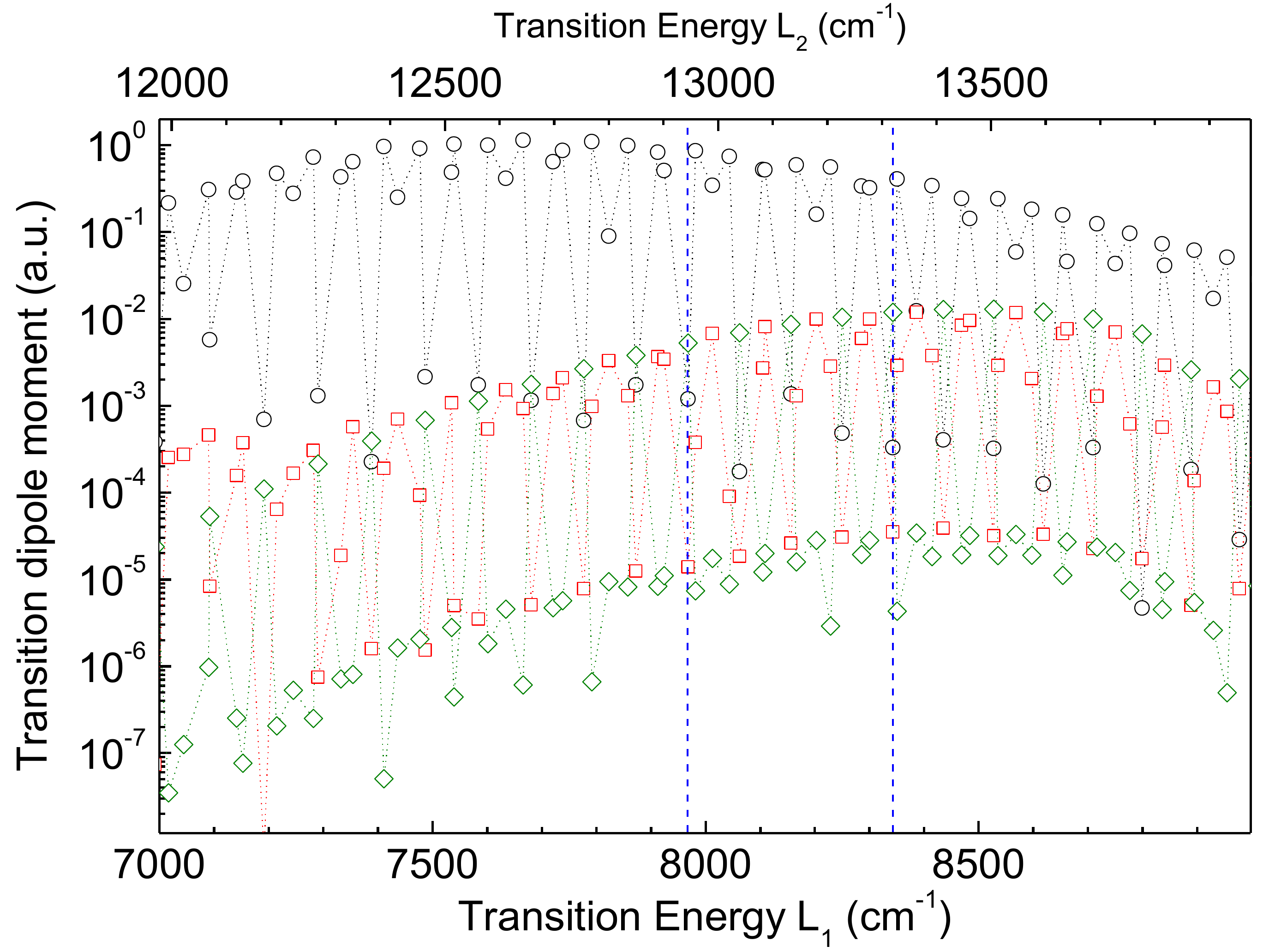}
	\caption {Computed absolute values of the transition dipole moments (TDMs) for the pump and dump transitions. The energy range represented by the dashed area in Fig.~\ref{fig1} and explored in the experiment is featured as vertical dashed lines. In the model (see text) the $\ket{1}$ state is the uppermost vibrational level $v_a = 21$ with $J_a = 0$ and $\ket{3} \equiv \ket{v_X = 0,J_X = 0}$. Open red squares (resp. green diamonds): TDM for the pump transition  driven by $L_1$ (lower horizontal energy scale) from level $\ket{1}$ toward $A-b_{0^+}$ (resp. $A-b_{1}$). Open black circles: TDM for the dump transition driven by $L_2$ (upper horizontal energy scale) from the $A-b_{0^+}$ component of level $\ket{2}$ to $\ket{3}$.}
	\label{fig2}
\end{figure} 

Following Ref.~\cite{docenko2007deperturbation}, we set up a four-coupled-channel hamiltonian \cite{borsalino2016} which accounts for the dominant SO interaction between the $A$ state and the $\Omega = 0^+$ component $b_{0^+}$ of the $b$ state, and for the weak rotational interactions with the other $b_1$ and $b_2$ components. Note however that the $b_2$ component is not relevant here, as we limit our study to the $J=1$ rotational levels of the excited state manifold. The $R$-dependent electronic transition dipole moments functions for the $X-A$ and $a-b$ transitions are those of our own quantum chemistry calculations \cite{aymar2007}. Vibrational energies are extracted from the diagonalization of the Hamiltonian matrix (see Eq.~2 of Ref.~\cite{borsalino2016}) with the mapped Fourier grid Hamiltonian (MFGH) method. The resulting transition dipole moments between levels $\ket{1}$, $\ket{2}$, and $\ket{3}$ are displayed in Fig.~\ref{fig2} for the levels $\ket{2}$ located in the energy range represented by the dashed area in Fig.~\ref{fig1}, which includes the experimental region. Note that they are labeled with a global index $v'$ referring to the eigenvalue order of the $A-b$ coupled system. We see that within the energy range explored in the experiment, the predicted TDMs have noticeable magnitude, allowing for well-balanced Rabi frequencies for the two $L_1$ and $L_2$ lasers.

\subsection{Overall spectrum}

\begin{figure}[t]
	\centering
	\includegraphics[width=0.4\textwidth]{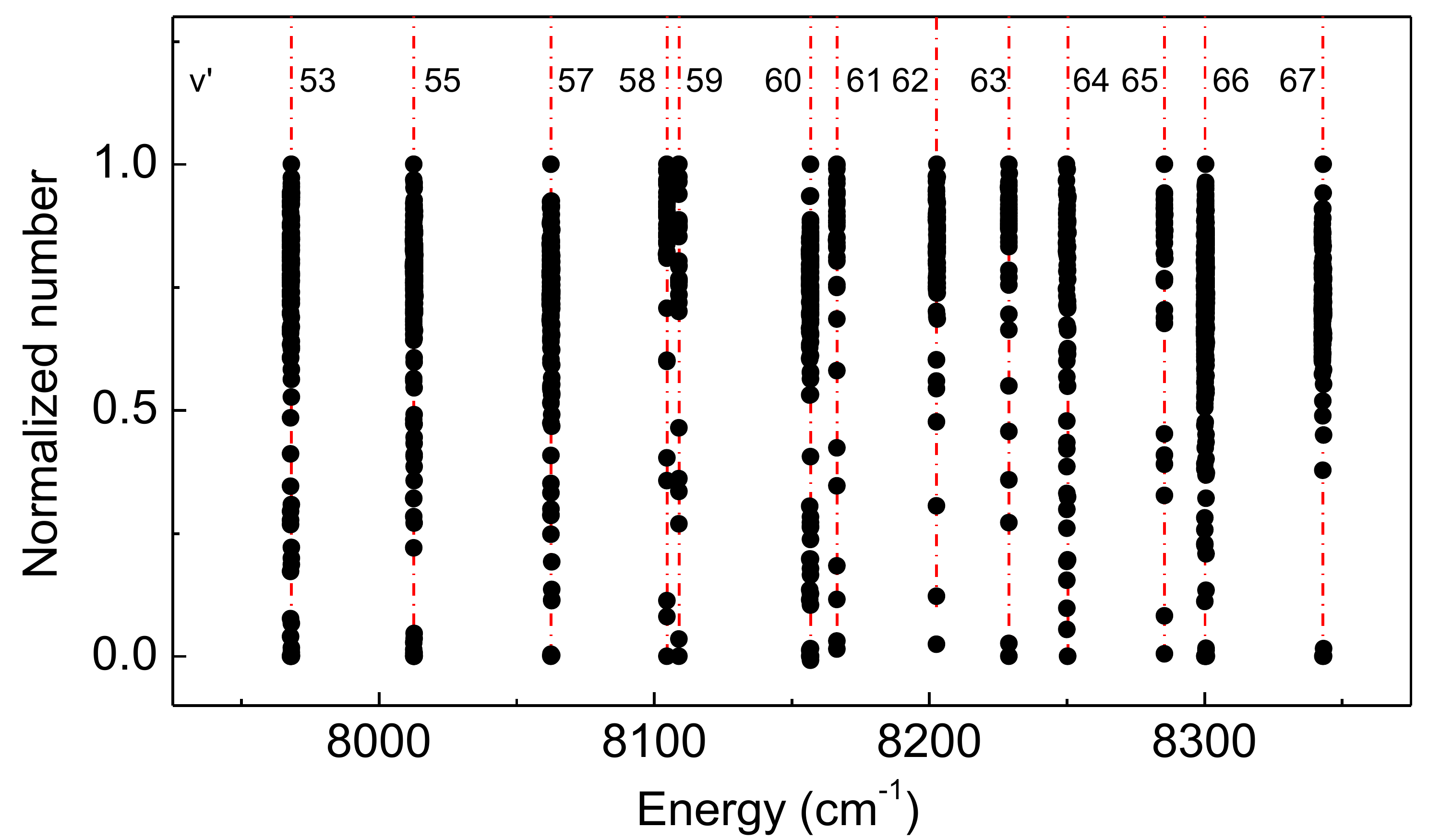}
	\caption{One-photon loss spectroscopy of Feshbach molecules for searching the intermediate level for STIRAP. Vibrational levels from $v'=53$ to $v'=67$ of the mixed $A-b_{0^+}-b_1$ system are observed. Note that the experimental signal for $v'=54,56$ was too weak to be observed.}
	\label{fig3}
\end{figure}

We report in Fig.~\ref{fig3} the lines detected in the range of $L_1$ wavenumbers from 7967~cm$^{-1}$ up to 8343~cm$^{-1}$. Our theoretical model indeed confirms, in agreement with Ref.~\cite{docenko2007deperturbation} that within this range the $A$ and $b$ states are strongly mixed and several levels have relatively large Franck-Condon overlaps with both $\ket{1}$ and $\ket{3}$ states. Each line of the spectrum corresponds to a vibrational level of the mixed $A-b_{0^+}-b_1$ states, and the corresponding assignments based on the above coupled-channel model are given in Table~\ref{components}. The observed levels are in good agreement with the calculations which confirms the high quality of the coupled potentials and the SO coupling functions. For each observed levels we also indicate the fractions of the $A$, $b_{0^+}$ and $b_1$ components of the $v'$ levels. Several of them are characterized by a significant component on both $A$ and $b_{0^+}$, expressing their strongly mixed singlet-triplet character. In contrast, due to the weakness of the rotational coupling of the $(A-b_{0^+})$ states to the $b_1$ state, levels with almost 100\% $b_1$ character exhibit a tiny fraction of $A$ state, which will be of crucial relevance in the following. 

Before proceeding to the analysis of the hyperfine structure of the intermediate excited levels, it is worthwhile to recall the basic selection rules governing the dipole-allowed transition in the present case. The parity with respect of an inversion in the laboratory frame of the total wave function of state $\ket{1}$  dominated by the $a^3\Sigma^+$ $\ket{v=21,J=1}$ level \cite{zhu2016long}, and of state $\ket{3} \equiv X^1\Sigma^+$ $\ket{v''=0,J''=0}$  is ``+'' for both. Thus the parity of the intermediate state $\ket{2}$ must be ``$-$''. For the levels with dominant $A-b_{0^+}$ character, the accessible rotational level is $J'=1$. For the levels with dominant $b_1$ character, both $J'=1$ and $2$ rotational states exhibit a ``$-$'' component. However, only $J'=1$ can couple to the $X^1\Sigma^+$ $\ket{v''=0,J''=0}$ level.

\begin{table}[t]
	\centering
	\caption{Assignment for the lines observed in Fig.~\ref{fig3}. Observed (Exp.) and calculated (Th.) vibrational level positions and their difference (Diff.) are displayed, with the squared component (expressed as percentages) of the calculated eigenvectors of the coupled $A-b_{0^+}-b_1$ system. For convenience purpose, large percentage values have been rounded off to the closest integer number, while those close to zero are explicitly displayed. Experimental values are measured on the coarse recorded spectra (with unresolved hfs) with an uncertainty of about 0.01~cm$^{-1}$.}
	\begin{tabular}{c|c|c|c|c|c|c}
		\hline\hline
		$v'$ & $A$ & $b_{0^+}$ & $b_1$ & Exp.       & Th.       & Diff. \\
		& (\%)& (\%)      & (\%)  & (cm$^{-1}$)&(cm$^{-1}$)&(cm$^{-1}$) \\
		\hline
		53 & 0.0003  & 0.0006  & 100     & 7967.84 & 7967.80 &0.04\\		
		54 & 97      & 3       & 0.0002  &   -     & 7981.77 & -\\	
		55 & 5       & 95      & 0.0006  & 8012.72 & 8012.75 &-0.03\\		
		56 & 95      & 5       & 0.0002  &   -     & 8043.69 & -\\		
		57 & 0.0001  & 0.0007  & 100     & 8062.46 & 8062.40 &0.06\\		
		58 & 77      & 23      & 0.0002  & 8104.66 & 8104.67 &-0.01\\		
		59 & 24      & 76      & 0.0005  & 8109.00 & 8109.02 &-0.02\\		
		60 & 0.0003  & 0.0007  & 100     & 8156.54 & 8156.48 &0.06\\		
		61 & 95      & 5       & 0.0002  & 8166.42 & 8166.43 &-0.01\\		
		62 & 6       & 94      & 0.0007  & 8202.64 & 8202.64 &0.00\\		
		63 & 97      & 3       & 0.000009& 8228.93 & 8228.91 &0.02\\		
		64 & 0.00003 & 0.0007  & 100     & 8250.04 & 8249.98 &0.06\\		
		65 & 64      & 36      & 0.0002  & 8285.50 & 8285.46 &0.04\\		
		66 & 38      & 62      & 0.0006  & 8300.15 & 8300.10 &0.05\\		
		67 & 0.00003 & 0.0007  & 100     & 8343.01 & 8342.89 &0.12\\		\hline\hline
	\end{tabular}	
	\label{components}
\end{table}

\section{Hyperfine structure of the excited levels}
\label{section3}

As shown in Table~\ref{components}, the observed vibrational levels have a dominant character either of $\Omega=0^+$ symmetry (from the coupled $A-b_{0^+}$ states) or $\Omega=1$ symmetry (from $b_1$). Their hyperfine structure is expected to be quite different, as discussed below.

The spectroscopic measurements are performed at the final value of the magnetic field 335.6~G. Thus only the projection $M_F$ on the quantization axis (along the magnetic field) of the total angular momentum $\textbf{F}=\textbf{J}+\textbf{I}_{\rm{Na}}+\textbf{I}_{\rm{Rb}}$ (associated to the quantum numbers $I_{\rm{Na}}=I_{\rm{Rb}}=3/2$, and $J$ is the rotational angular momentum of the molecule)  is a good quantum number. The $\ket{1}$ state has $M_F=2$ \cite{zhu2016long}, allowing transitions towards $\ket{2}$ levels with $M_F'=1,3$ (resp. $M_F'=2$) when $L_1$ has a $\sigma^{\pm}$ (resp. $\pi$ polarization). At 335.6~G, the lowest level of the $X^1\Sigma^+$ $\ket{v''=0,J''=0}$ manifold has $M''_F=3$. Therefore various polarization combinations of $L_1$ and $L_2$ are possible to achieve the STIRAP transfer, provided that the chosen $\ket{2}$ level fulfills the required selection rules.

\subsection{The $b_1$ levels}

\begin{figure}[!t]
	\includegraphics[width=0.4\textwidth]{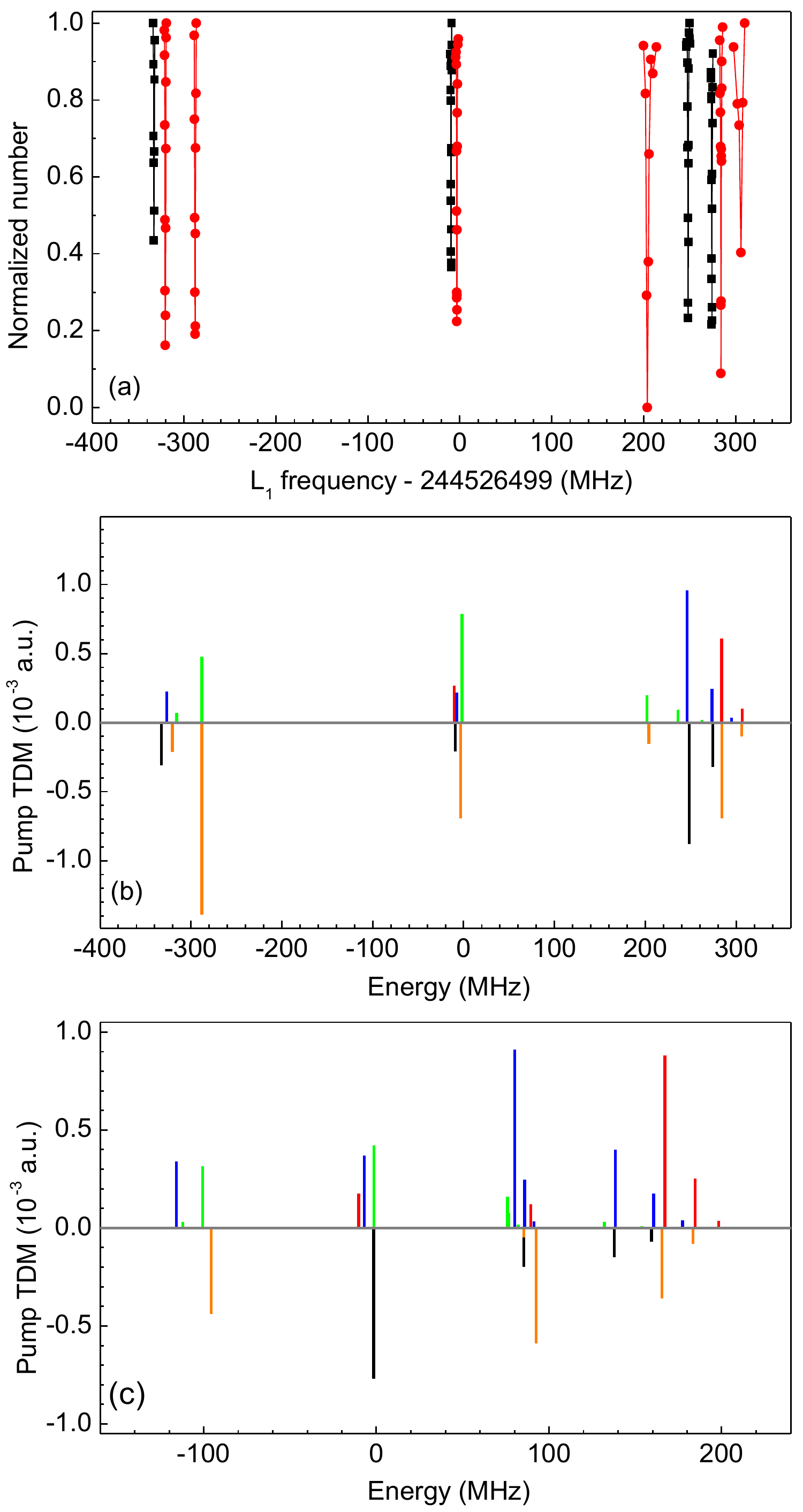}
	\caption{(a) High resolution spectrum recorded at 335.6~G of (a) the $\ket{v'=60,J'=1}$ level (with dominant $b_1$ character) exhibiting a large and fully-resolved hyperfine structure. The data in black points are obtained with vertical polarized light ($\pi$ transitions), while the red squares are obtained with horizontal polarization ($\sigma$ transitions). Being a piecewise spectrum, the intensities of the lines are arbitrarily normalized from one piece to the other. (b) Comparison between the experimentally extracted TDMs (lower panel) with those yielded by the perturbative model (upper panel). Lines are represented as vertical bars as follows. Upper panel: for $\sigma$ transitions (with $M'_F$=3 in red and and $M'_F=1$ in green), and $\pi$ transitions (with $M'_F$=2 in blue); lower panel: for $\sigma$ transitions in orange, and for $\pi$ transitions in black. (c): same as (b) for the $\ket{v'=60,J'=2}$ level with zero energy corresponding to $L_1$ frequency of 244.534386~THz.}
	\label{fig4}
\end{figure}

Figure~\ref{fig4}(a) shows a typical detailed spectrum of the $\ket{v'=60,J'=1}$ level which was our original choice for STIRAP. The hyperfine coupling is strong enough to induce significant splitting between different sub-levels. The whole spectrum spans more than 600~MHz, nearly 60 times larger than the one observed for the $\ket{v'=59,J'=1}$ level (Fig.~\ref{fig5}). However, even with the position of the $X^1\Sigma^+$ $\ket{v''=0,J''=0}$ level known from the spectroscopy using the $\ket{v'=59,J'=1}$ level (see next section) and the highest $L_2$ power available, no two-photon resonance was ever observed with this $\ket{v'=60,J'=1}$ level. This is consistent with the results reported in Table~\ref{components} where the $\ket{v'=60}$ level is predicted with a very little $A^1\Sigma^+$ character, at the limit of what could be used for an efficient transfer to the absolute ground state \cite{debatin2011molecular}.

The hyperfine structure of the $\ket{v'=60}$ level can be modeled with the help of an effective Hamiltonian $\textbf{H}^{\rm eff}$ inspired from the asymptotic model of Ref.~\cite{orban2015model}, expressing the interaction between the molecular orbital angular momentum $\textbf{L}$ and spin $\textbf{S}$ of the molecule with the magnetic field $\textbf{B}$, and with the individual nuclear spins $\textbf{I}_{\rm{Na}}$ (with projection $M'_{\rm{Na}}$) and $\textbf{I}_{\rm{Rb}}$ (with projection $M'_{\rm{Rb}}$) through the first-order Hamiltonian $\textbf{H}^{(1)}_{\rm hfs}$. The interaction with the nuclear quadrupole moment is neglected. We choose the Hund's case (c) spin-decoupled basis $\ket{\alpha} \equiv \ket{J'\Omega M'_J;I_{\rm{Na}}M'_{\rm{Na}};I_{\rm{Rb}}M'_{\rm{Rb}}}$ with 48 vectors. The $\textbf{H}^{\rm eff}$ matrix elements are obtained in a perturbative way relative to the bare energy $E_{v'J'}$of the excited $\ket{2}$ level:
\begin{equation}
\label{Ham}
	\begin{split}
	H^{\rm eff}_{\alpha' \alpha} = E_{v'J'}+w_{\rm{Na}}I_{\rm{Na}}H^{(1)}_{{\rm{hfs}}, \alpha' \alpha}({\rm{Na}}) + w_{\rm{Rb}}I_{\rm{Rb}}H^{(1)}_{{\rm{hfs}}, \alpha' \alpha}({\rm{Rb}})\\
	+ g_S\mu_N(\textbf{S}\cdot\textbf{B})_{\alpha' \alpha} + g_L\mu_N(\textbf{L}\cdot\textbf{B})_{\alpha' \alpha} + \left(\frac{1}{2\mu R^2}\textbf{O}^2 \right)_{\alpha' \alpha},
	\end{split}
\end{equation}
where the notation $()_{\alpha' \alpha}$ holds for matrix elements of composite operators. The derivation of these matrix elements is described in the Appendix. Equation~(\ref{Ham}) assumes that the couplings between different electronic, vibrational and rotational levels are negligible: $J'$ and $\Omega$ remain good quantum numbers, while the same vibrational wave function (with energy $E_{v'J'}$) is taken for all sublevels. The other quantities are: the spin g-factor $g_S$, the electron orbital g-factor $g_L$, the Bohr magneton $\mu_B$, the mechanical molecular rotation $\textbf{O}$, and the effective coupling constants $w_{\rm{Na}}$ and $w_{\rm{Rb}}$. Among the 48 vectors, only 9 $M'_F=1$, 6 $M'_F=2$ and 3 $M'_F=3$ sublevels are experimentally accessible starting from the $\ket{1}$ state with $M_F=2$. We note that these numbers cannot be trivially derived, and require a careful accounting of all the 48 levels of the manifold. All these levels are doubly-degenerate as $\Omega=1$, and this degeneracy is in principle lifted by the electronic quadrupole interaction which is not included here.

Based on this model, a fitting procedure was performed to reach the best possible matching between the calculated eigenenergies of the effective Hamiltonian [Eq.~(\ref{Ham})] and the observed energies. As shown in Fig.~\ref{fig4}(b), a good agreement is obtained by adjusting the two hyperfine coupling constants only, leading to $w_{\rm{Na}}=23.9295$~MHz and $w_{\rm{Rb}}=86.2171$~MHz. Moreover, although the transition strengths are not included in the fitting procedure, the theoretically predicted ones are consistent with experimentally extracted values (see Sec.~\ref{subsection2}). We further applied this model with the above coupling constants to the $\ket{v'=60,J'=2}$ level, yielding again a good overall agreement with the experimental measurements of the energy positions and intensities [Fig.~\ref{fig4}(c)].

\subsection{The $A-b_{0^+}$ levels}

\begin{figure}[!t]
	\includegraphics[width=0.4\textwidth]{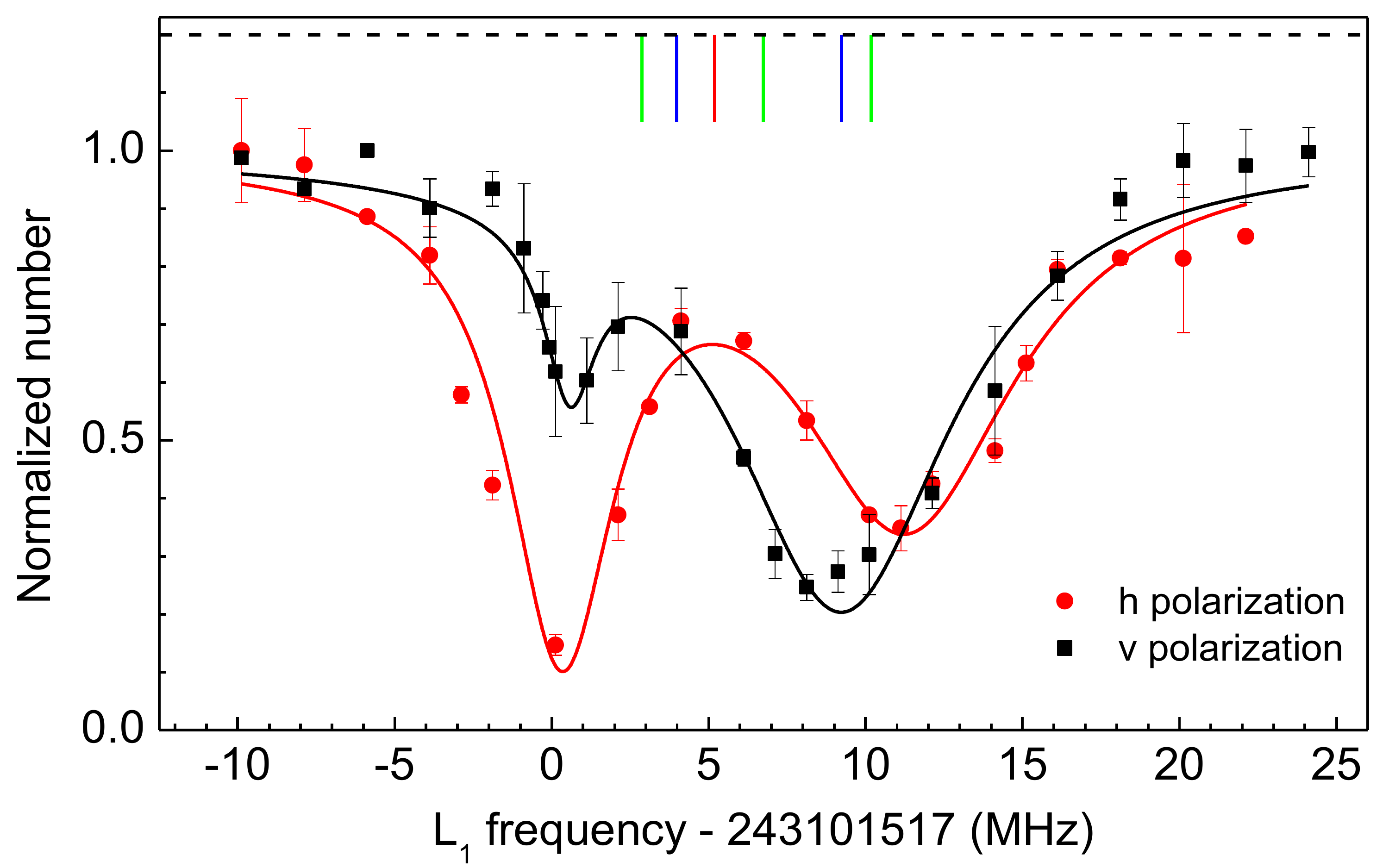}
	\caption{High resolution spectrum recorded at 335.6~G of the $\ket{v'=59,J'=1}$ level (with dominant $A-b_{0^+}$ character) with partially-resolved hyperfine structure. Black points: vertically (v) polarized light ($\pi$ transitions). Red squares: horizontally (h) polarized light ($\sigma$ transitions). The vertical bars indicate line positions from the theoretical model with the same color convention than in Fig.~\ref{fig4}.}
	\label{fig5}
\end{figure}

Figure~\ref{fig5} exemplifies the expected much smaller hyperfine structure of the $0^+$ dominated levels. Only two loss peaks are observed and the splitting between them, which depends on the polarization of $L_1$, is about 10~MHz. The corresponding level $\ket{v'=59,J'=1}$ is an admixture of 24$\%$ in $A^1\Sigma^+_{0^+}$ and 76$\%$ in $b^3\Pi_{0^+}$. The calculation indicates that the transition strength between this level and the $X^1\Sigma^+$ $\ket{v''=0,J''=0}$ one is indeed strong. Actually, our initial successful search of the $X^1\Sigma^+$ $\ket{v''=0,J''=0}$ level with two-photon spectroscopy was conducted with this intermediate level. However, the  population transfer with this route failed due to the unresolved hyperfine structure, as discussed in the introduction.

In principle, the effective Hamiltonian model of Eq.~(\ref{Ham}) could also be applied to the $\ket{v'=59}$ level. However, all the terms in this equation cancel for $\Omega=0$, so that second-order terms should be introduced, like the interaction involving the nuclear quadrupole moment, or the interaction with neighboring rovibrational levels. These are all small competing quantities which would be described with a significant number of parameters regarding the number of measured lines leading to more complicated calculations than in the previous case. The effort is probably unworthy as the hyperfine structure cannot be fully resolved experimentally. But it is worthwhile to check the predictive power of the asymptotic model reported in Ref.~\cite{orban2015model}, where the hyperfine structure is included as a perturbation to the Hund's case (c) PECs without rotation, and parametrized by the atomic hyperfine structure. A manifold of $(2I_{\rm Na}+1)(2I_{\rm Rb}+1)=16$ $0^+$ PECs is obtained, among which one $M'_F=3$, two $M'_F=2$, and three $M'_F=1$ are accessible from the $\ket{1}$ level (with $M_F=2$). The vibrational energies are deduced by averaging these PECs over the vibrational wave function of the $\ket{v'=59}$ level resulting from a MFGH calculation performed on the coupled $A-b$ system. These values are reported as colored ticks in Fig.~\ref{fig5}. Given the simplicity of the model, these individual energies should not be considered as rigorous. However, it is noticeable that a structure spread over over 7~MHz is found, in rather good agreement with the spread of the experimental structure. The asymptotic model seems to provide a reasonable representation of the strength of the molecular hyperfine structure. Actually, we also used this asymptotic model for the $(\Omega=1)$~-dominated level $\ket{v'=60}$, which allowed us to derive initial values of the coupling constants $w_{\rm{N}a}=8.7$~MHz and $w_{\rm{R}b}=87$~MHz, which happened to be consistent with the fitted values.

\subsection{Hyperfine structure induced by accidental $0^+$ and $0^-$ mixing}

Most of the other vibrational levels we observed have similar structures. Actually, following the above discussions on the two types of ``normal'' hyperfine structures, it seems from Table~\ref{components} that none of the observed vibrational levels can simultaneously satisfy all requirements for an efficient STIRAP. Either they have no $\Omega=1$ component and thus have unresolvable hyperfine structures (Fig.~\ref{fig5}), or they have not enough singlet character to ensure the efficient radiative coupling to the absolute ground state (Fig.~\ref{fig4}). Thus those levels are not suitable for STIRAP. 

\begin{figure}[!t]
\includegraphics[width=0.4\textwidth]{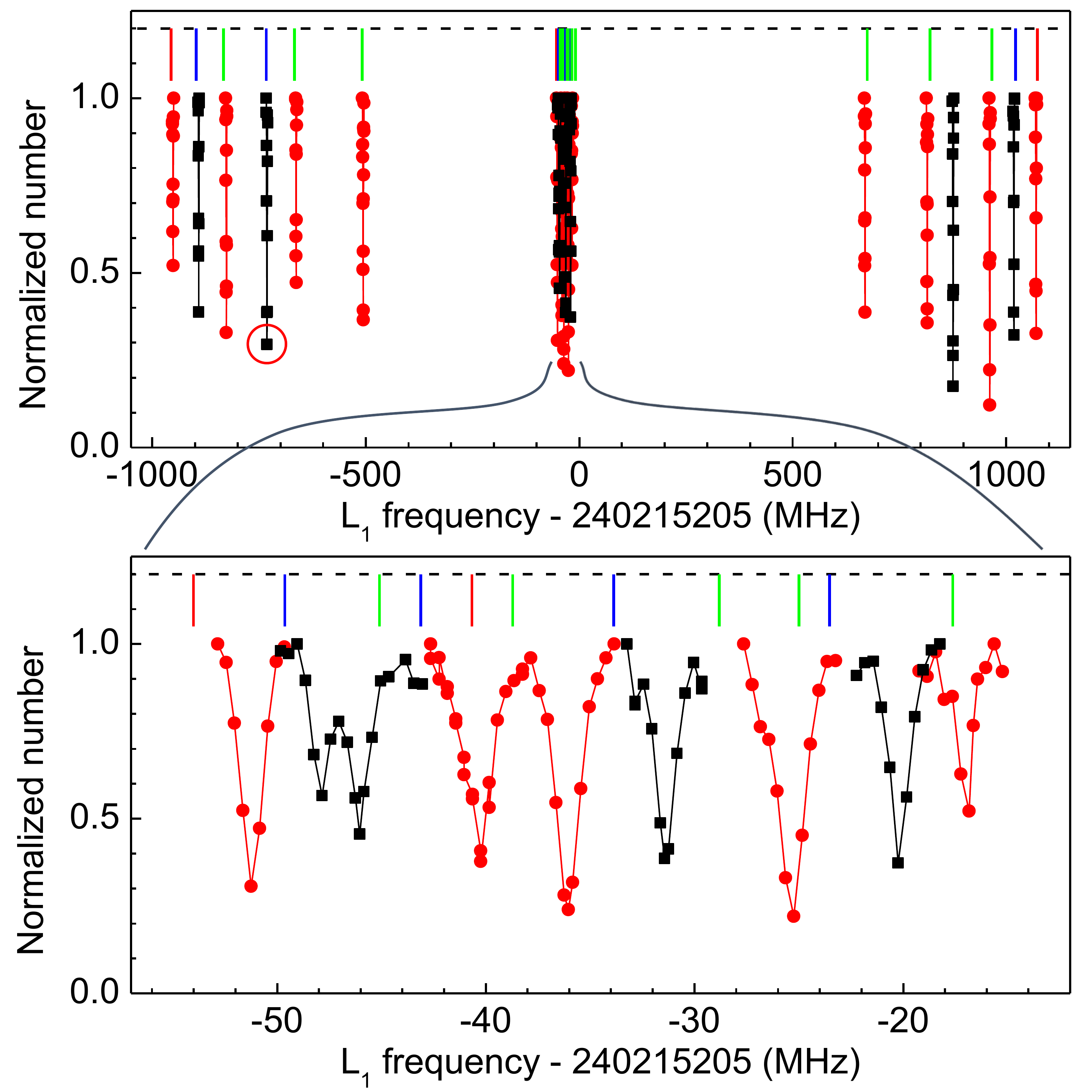}
	\caption{(a) High-resolution spectrum (recorded at 335.6~G) showing 21 resolved lines for the hyperfine structure of the $\ket{v'=55,J'=1}$ level. Black points: vertical polarized light ($\pi$ transitions). Red squares: horizontal polarization ($\sigma$ transitions). The vertical bars indicate line positions from the theoretical model with the same color convention as in Fig.~\ref{fig4}. The red open circle marks the hyperfine level used in STIRAP.  (b) Blow-up of the central group of lines. The bars indicate the position of the calculated levels, with the color code: for $\sigma$ transitions (with $M'_F$=3 in red and $M'_F=1$ in green), and $\pi$ transitions (with $M'_F$=2 in blue).}
	\label{fig6}
\end{figure}

Fortunately enough, we recorded a well-resolved set of lines around the $v'=55$ level, spanning a spectral range of about 2~GHz (Fig.~\ref{fig6}), much larger than any expected hyperfine structure. It is composed of a central band of about 35~MHz wide, and two approximately symmetric bands 800~MHz away. Our analysis below assigns these unexpected structures to the accidental coupling between two quasi-degenerate levels of the $A-b_{0^+}$ system and of the $c_{0^-}-b_{0^-}$ system (Fig.~\ref{fig1}). The $v'= 55$ level turns out to be a great choice for STIRAP. Choosing the appropriate line among the structure as level $\ket{2}$, a population transfer efficiency over 95$\%$ toward level $\ket{3}$ has been achieved.

In $^{87}$Rb$_2$ \cite{Deiss2015mixing} it has been shown that the coupling between $\Omega=0^+$ and $\Omega=0^-$ states can induce an unusually large hyperfine structure of several hundreds of MHz. Thus we examined the possibility for a similar pattern in NaRb. In $^{23}$Na$^{87}$Rb, the energies of the vibrational levels of the $\Omega=0^+$, $\Omega=1$ and $\Omega=2$ components of the $b$ state are well-known thanks to the deperturbation analysis of Ref.~\cite{docenko2007deperturbation}. However, the spectroscopy of the nearby $\Omega=0^-$ state (resulting from the weak perturbation of the $b^3\Pi$ state by the $c^3\Sigma^+$ state) is still unknown, as pure triplet levels are not easily accessible from the singlet ground-state molecule.  

We build our model starting from the energies of the $A-b_{0^+}$ levels of Ref.~\cite{docenko2007deperturbation}. To determine the energy levels of the $c_{0^-}-b_{0^-}$ system, we used the deperturbed $b^3\Pi$ PEC, while the $c^3\Sigma^+$ PEC comes from our quantum chemistry calculations \cite{aymar2007calculations}. The diagonal elements of the SO coupling matrix are taken from Ref.~\cite{docenko2007deperturbation}, while the unknown off-diagonal term is approximated with the $R$-dependent \textit{ab initio} data computed in Ref.~\cite{Kotochigova2009} for KRb. For comparison, we also performed the same calculation with an $R$-independent off-diagonal coupling (set to the atomic value).

 \begin{figure}[!t]
 	\centering
 	\includegraphics[width=0.4\textwidth]{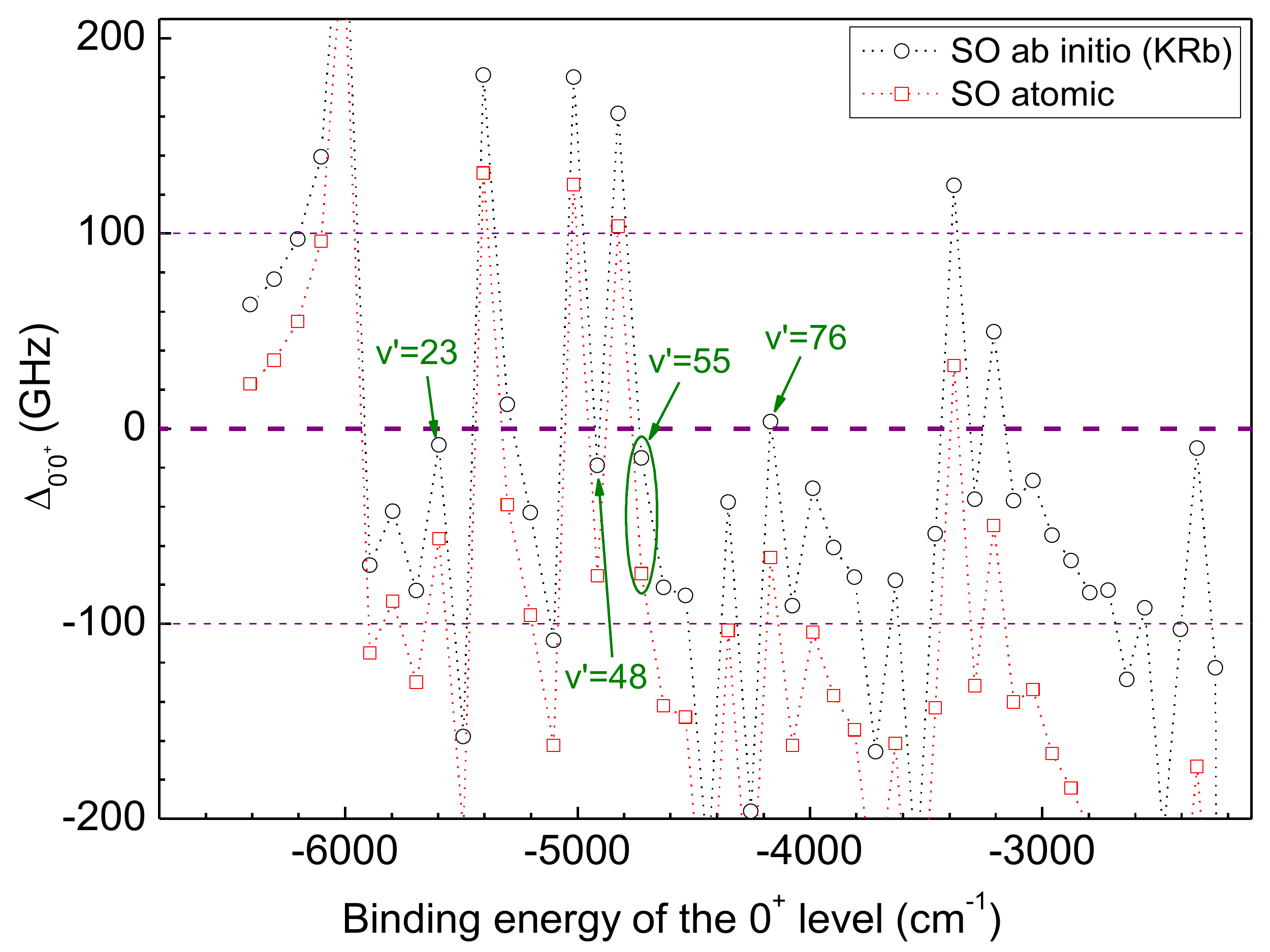}
 	\caption{Computed values of the energy spacings $\Delta$ between the vibrational levels of $0^+,v'_{0^+},J'=1$ with a dominant $b$ character and the corresponding  $0^-,v'_{0^-},J'=0$  levels. Calculations are done with an \textit{ab initio} $R$-dependent SO coupling \cite{Kotochigova2009} (black circles) and with an $R$-independent (atomic) one. The $v'=55$ level is circled out as the one which is used for STIRAP.}
 	\label{fig7}
 \end{figure}

It is well-established that the PECs for the $b_{0^+}$ and $b_{0^-}$ components are close to each other (see for instance the \textit{ab initio} data of Ref.~\cite{korek2008b} showing that their minimum is separated by less than 5~cm$^{-1}$ from each other). In order to evaluate the chance for two levels with dominant $b^3\Pi$ character to be close to each other in the $A-b_{0^+}$ and the $c_{0^-}-b_{0^-}$ systems, we represent in Fig.~\ref{fig7} the energy spacings $\Delta_{0^-0^+}$ between  $0^+,v'_{0^+},J'=1$ levels and their $0^-,v'_{0^-},J'=0$ neighboring levels (both with dominant $b^3\Pi$ character). The choice for $J'$ values will become clear later on. If all SO couplings are set to zero, all pairs of $0^+$ and $0^-$ neighboring levels are obviously degenerate ($\Delta=0$) as they belong to the $b^3\Pi$ state. As the $b^3\Pi$ and $c^3\Sigma^+$ PECs are quite far apart from each other in the energy range of interest (namely the potential well of the $b^3\Pi$ state), the $c$ state only slightly perturbs the $b$ vibrational progression so that the resulting $b_{0^-}$ vibrational progression remains quite regular. In contrast the strong SO coupling in the  $0^+$ case (induced by the short-range crossing between the $b^3\Pi$ and $A^1\Sigma^+$ PECs) results in a strongly perturbed $A-b_{0^+}$ vibrational progression. This yields $\Delta$ values heavily varying between -200~GHz to +200~GHz. For the $v'=55$ level, we computed $\Delta \approx15$~GHz with the $R$-dependent SO coupling, but a slight change in the parametrization of the SO coupling could significantly reduce $\Delta_{0^-0^+}$. We note that in this model computation, several pairs of levels are found close to each other (Fig.~\ref{fig7}) which suggests that indeed the invoked reason for explaining the $v'=55$ hyperfine structure is sensible.

We have therefore extended our basis used in the effective Hamiltonian model [Eq.~(\ref{Ham})] to include both $\Omega=0^+$ and $\Omega=0^-$. We thus have the basis $\ket{J'\Omega\epsilon M_{J'};I_{\rm{Na}}M'_{\rm{Na}};I_{\rm{Rb}}M'_{\rm{Rb}}}$, with $\epsilon=\pm1$ the parity of the electronic wavefunction and $J'$ varying from 0 to 2. As the total parity is conserved, we only considered basis vectors with a total ``$-$'' parity (the one that can be reached experimentally starting from a Feshbach level of ``+'' parity). This implies restricting the basis set to $0^+$ (resp. $0^-$) levels with odd (resp. even) $J'$ values. We have finally 144 basis vectors $\ket{\alpha}$ distributed over the 11 possible $M'_F$ values. This means for the experimentally reachable projection : 25 sublevels with $M'_F=1$, 18 with $M'_F=2$, and 10 with $M'_F=3$. The matrix elements of the effective Hamiltonian are
\begin{equation}
\label{Ham3}
\begin{split}
 H^{\rm eff}_{\alpha' \alpha} = \delta_{\epsilon\epsilon'}E_{\epsilon} &+ w_{\rm{Na}} I_{\rm{Na}} H^{(1)}_{\rm hfs, \alpha' \alpha}(\rm{Na}) \\
+ w_{\rm{Rb}} I_{\rm{Rb}} H^{(1)}_{\rm hfs, \alpha' \alpha}(\rm{Rb}) &+ \frac{1}{2}eqQ_{\rm{Na}} H^{(2)}_{\rm hfs, \alpha' \alpha}(\rm{Na})\\
+ \frac{1}{2}eqQ_{\rm{Rb}}H^{(2)}_{\rm hfs, \alpha' \alpha}(\rm{Rb}) &+ w_{z} g_S\mu_N (\textbf{S} \cdot \textbf{B})_{\alpha' \alpha} \\
+ w_{z} g_L\mu_N (\textbf{L}\cdot \textbf{B})_{\alpha' \alpha} &+ \left(\frac{1}{2\mu R^2}\textbf{O}^2 \right)_{\alpha' \alpha}.
\end{split}
\end{equation}
We note $E_{\epsilon}$ the purely vibrational energies: $E_{+}$ (resp. $E_{-}$) for the $0^+$ (resp. $0^-$)level. The rotational energies relevant for the present basis set are set as $E_{+}+2B_v^{0^+}$ (for $J'=1$), and $E_{-}+6B_v^{0^-}$ (for $J'=2$), involving rotational constants $B_v^{0^+}$ and $B_v^{0^-}$. Thus we have the identity $E_{-}=E_{+}+2B_v^{0^+}+\Delta_{0^-0^+}$. The effective Hamiltonian now includes the second-order operators $H^{(2)}_{\rm hfs}$ for the hyperfine interaction with the nuclear quadrupole moments $eqQ_{\rm{Na}}/2$ and $eqQ_{\rm{Rb}}/2$ (where $eqQ$ is the conventional notation for this quantity \cite{broyer1978}). The constant $w_z$ accounts for the fact that this coupled state has not a pure spin character, and rather an unknown singlet-triplet mixture, thus influencing the Zeeman energy. The derivation of these matrix elements is described in the Appendix.

\begin{figure}[!t]
 	\centering
	\includegraphics[width=0.4\textwidth]{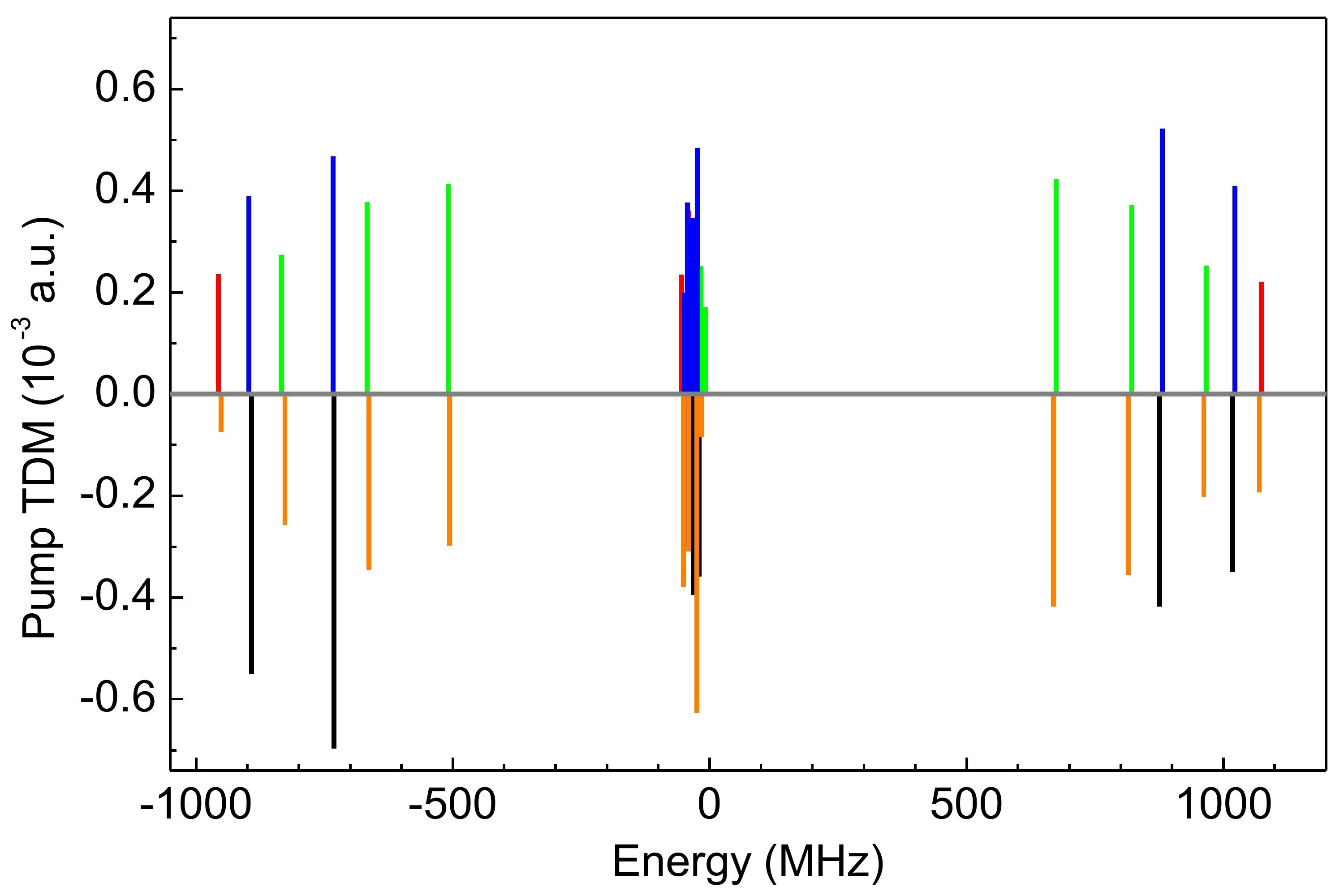}
     \caption{ Computed (upper panel) and observed (lower panel) spectra for the $\ket{v'=55,J=1}$ level. The zero energy corresponds to the level energy position without Zeeman and/or hfs. Upper panel: for $\sigma$ transitions (with $M'_F$=3 in red and and $M'_F=1$ in green), and $\pi$ transitions (with $M'_F$=2 in blue); lower panel: for $\sigma$ transitions in orange, and for $\pi$ transitions in black. The experimental intensities are calibrated following Sec.~\ref{subsection2}.}
		\label{fig8}
\end{figure}

Due to their expected small magnitude, the $H^{(2)}_{\rm hfs}$ terms will mostly concern the structure of the central band. Therefore, in a first step we set $eqQ_{\rm{Na}}/2$ and $eqQ_{\rm{Rb}}/2$ to zero,and $w_{z}=0.94624789$ is the computed percentage of $b_{0^+}$ character of the $v'=55$ (namely, the value which was rounded off in Table~\ref{components}). Four parameters are left free to achieve a rough fit of the broad structures: $w_{\textrm{Na}}$, $w_{\textrm{Rb}}$, the splitting $\Delta_{0^-0^+}$, and the energy $E_{+}$. The quantity $E_{+}+2B_{0^+}$ is initially fixed to the tabulated energy of the $\ket{v'=55, J'=1}$ level (with the experimental value $B_{0^+}=1946.054947$~MHz extracted from Ref.~\cite{docenko2007deperturbation}), but as the center-of-gravity of the structure is \textit{a priori} unknown, $E_{+}$ must be considered as a variable parameter. Thus we obtain initial values to perform a second fit now including $w_z$ and $B_{0^-}$. As illustrated in Fig.~\ref{fig8}, a satisfactory agreement is obtained with the final set of parameters: $w_{Na}=245.291$~MHz, $w_{Rb}=520.75$~MHz, $\Delta=181.228$~MHz, $w_{z}=0.952796$, $B_v^{0^-}=1960.6$~MHz. We see that $B_v^{0^-}$ and $B_v^{0^+}$ are very close to each other (consistent with the closeness of their PECs), and that $w_z$ is only slightly changed by the fit (so that the computed triplet-singlet mixture seems accurate). 

In an attempt to represent the central lines, all the previous parameters are kept fixed to their values above, while the coupling constants of the electric quadrupole interaction are varied, yielding $\frac{1}{2}eqQ_{Na}=-11.4457$~MHz, $\frac{1}{2}eqQ_{Rb}=6.9642$~MHz. The resulting line positions are reported in Fig.~\ref{fig6}(b): just like for the previous $(A-b_{0^+})$ case, the width of the manifold is in satisfactory agreement with the recorded lines, but the positions cannot be accurately reproduced as several small competing interactions that could bring a contribution at the MHz level have been neglected. We point out though that adding several competing operators will increase the complexity of the model (\textit{i.e.} the number of free parameters) and it will become difficult to derive meaningful results with so few experimental data points being available for the regression procedure. Uncertainties on the exact position of the sublines will also make difficult any attempt to improve the current model. 

\subsection{Calibration of the pump transition strength for STIRAP}
\label{subsection2}

\begin{figure}[!t]
	\centering
	\includegraphics[width=0.45\textwidth]{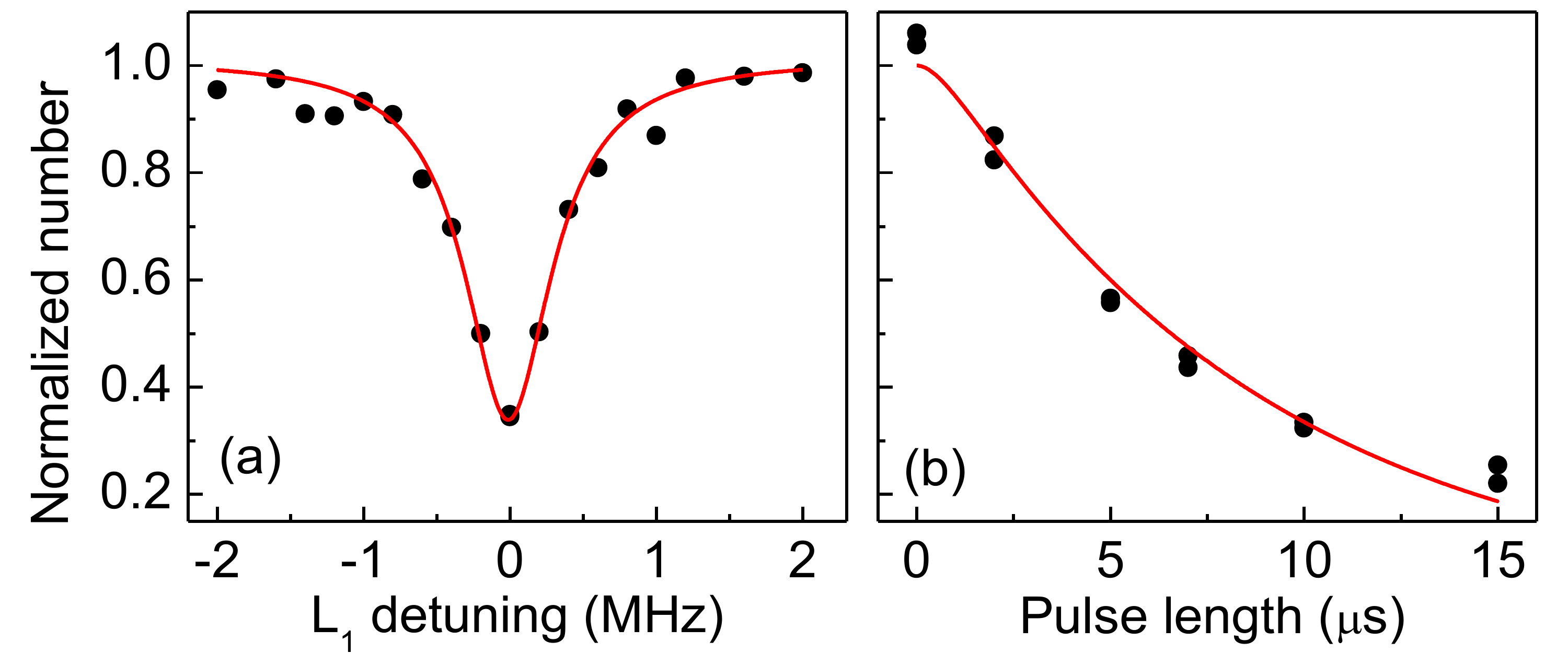}
	\caption{Pump transition strength calibration from the Feshbach state $\ket{1}$ to the selected hyperfine level $\ket{2}$ in Fig.~\ref{fig6}, based on the measurement of the variations of the population of Feshbach molecules when the $L_1$ laser is on with a power of 0.456~mW and a waist of about 45~${\mu}$m. (a) Line shape obtained with a $L_1$ pulse length of 10~$\mu$s. (b) Time evolution of Feshbach molecules with $L_1$ on resonance. The red curves result from the simultaneous fit of the two data sets to extract the TDM and the excited state lifetime. }
	\label{fig9}
\end{figure}

To calibrate the coupling strength of the pump transition from the Feshbach state $\ket{1}$ to the selected sub-structure $\ket{2}$ of $\ket{v'=55,J'=1}$ (Fig.~\ref{fig6}), we scan the frequency of the laser $L_1$ while keeping the pulse duration fixed [Fig.~\ref{fig9}(a)] and the pulse duration of the laser $L_1$ while keeping its frequency on resonance [Fig.~\ref{fig9}(b)]. The resulting line shape and time evolution can be fitted following a standard procedure~\cite{debatin2011molecular}. The excited-state spontaneous decay rate is found to be $\gamma=2\pi\times 0.67$~MHz, corresponding to a lifetime of 238~ns. Together with the measured beam profile of $L_1$, the normalized Rabi frequency is determined as $\bar{\Omega}_1=2\pi\times 1.01 \rm kHz\times\sqrt{I/(mW/cm^2)}$, which corresponds to a TDM of about 0.0007~a.u.. Such a transition strength allows for generating Rabi frequencies larger than  1~MHz with our available laser power. This transition strength coupled with the large hyperfine splitting make the $\ket{v'=55,J'=1}$ level an ideal intermediate level for efficient STIRAP. 

For completeness, we report in the Appendix the complete list of the sublevels of the $\ket{v'=55,J'=1}$ manifold, with their main experimental and theoretical characteristics. In particular, it is found that the TDM extracted from experiment for the chosen $\ket{2}$ level is consistent with the calculated one equal to 0.00047~a.u.. In contrast, the lifetime of the selected hyperfine level is calculated at 531~ns, twice larger than the measured value above. We argue in the Appendix about the possible explanation: the accidental predissociation of the $(A-b_{0^+})$ induced by the rotational coupling of the $b$ PEC with the repulsive branch of the the $a^3\Sigma^+$ state at short distances (see Fig.~\ref{fig1}), as already suggested in Ref.~\cite{schmidt-mink1985b}.
 
\section{Ground-state spectroscopy}
\label{section4}

\subsection{Rotational and hyperfine structures of the $v''=0, J''=0$ ground-state level}

\begin{figure}[!t]
	\centering
	\includegraphics[width=0.4\textwidth]{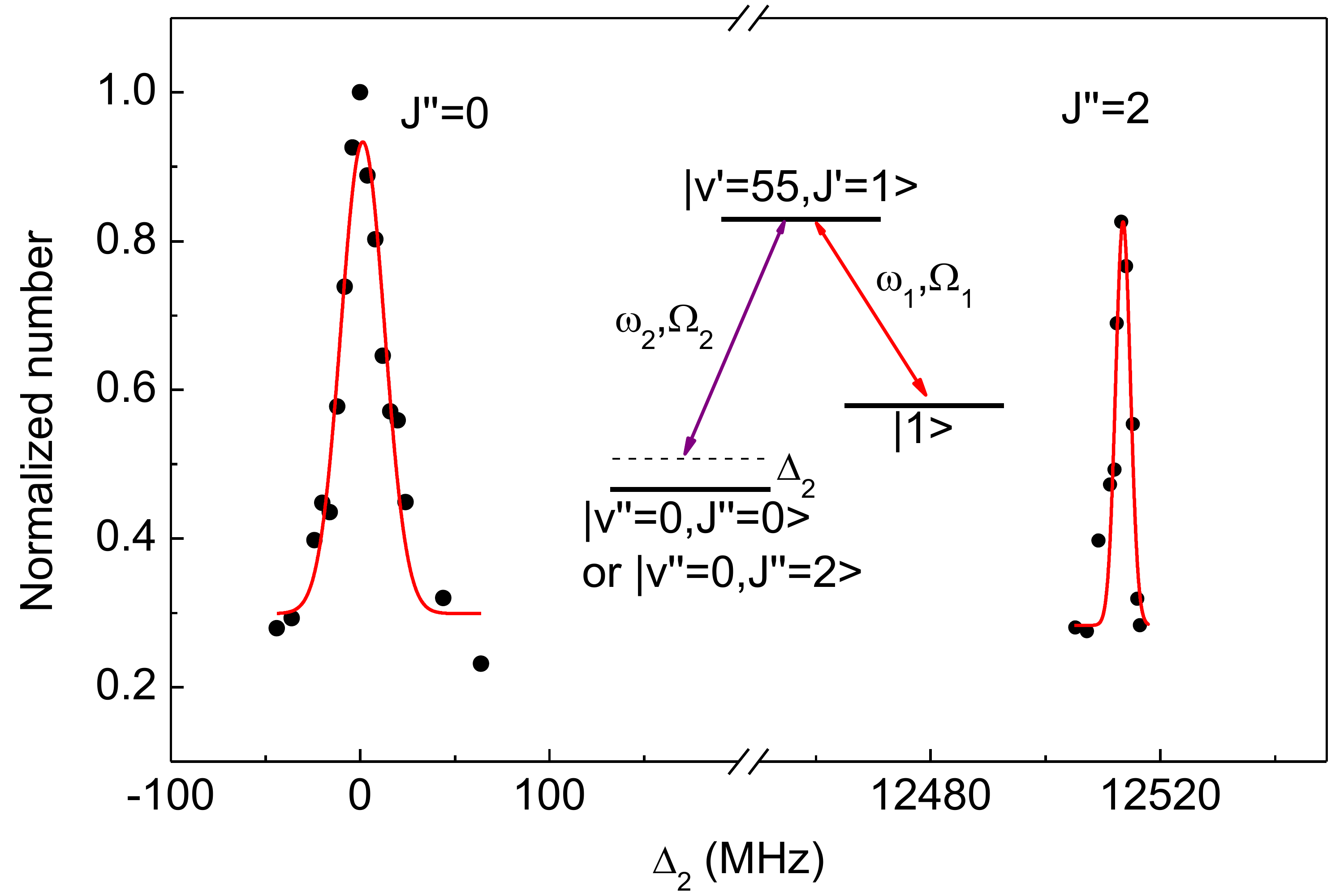}
	\caption{Two-photon spectroscopy of the $\ket{v''=0,J''=0}$ and $\ket{v''=0,J''=2}$ ground state levels via the $\ket{v'=55,J'=1}$ excited intermediate level. The scheme indicates that $L_1$ is kept on resonance, while the frequency of $L_2$ (expressed as the detuning $\Delta_2$ with respect to the resonance). A much larger dump transition Rabi frequency (induced by a laser power as high as 27699.9~mW/cm$^2$) than the pump one is used for obtaining this spectrum, so that the hyperfine structure is not resolved. The red curves are Gaussian fits for extracting the line centers. }
	\label{fig10}
\end{figure}

Because of its $A^1\Sigma^+$ character as high as 5$\%$, the $\ket{v'=55,J'=1}$ levels can efficiently couple to the $\ket{v''=0,J''=0}$ level of the $X^1\Sigma^+$ state. Figure~\ref{fig10} displays the two-photon spectrum obtained by scanning the dump laser $L_2$ with $L_1$ kept on resonance. When $L_2$ is tuned to the resonance between the $\ket{2}$ level and the $\ket{v''=0}$ level of the $X^1\Sigma^+$ state (the $\ket{3}$ level), the position of the excited state is shifted due to the AC Stark effect. The $L_2$ laser can be brought to power large enough to shift $L_1$ off resonance. As a result, the loss of Feshbach molecules caused by $L_1$ is reduced and a two-photon resonance shows up as a recovery of Feshbach molecules. Figure~\ref{fig10} includes two such two-photon resonances to the $\ket{v''=0,J''=0}$ and $\ket{v''=0,J''=2}$ levels. From the interval of the two peaks, one can extract the related rotational constant $B_0^{X}=0.0697$~cm$^{-1}$, consistent with previous result~\cite{wang2013observation}. From the frequencies of $L_1$ and $L_2$, the binding energy of the absolute ground state of $^{23}$Na$^{87}$Rb relative to the origin given by the hyperfine center-of-mass of both atoms at zero magnetic field is determined to be $D_0^{X}=4977.308(3)$~cm$^{-1}$, compared to the previous determinations $D_0^{X}=4977.536$~cm$^{-1}$ \cite{docenko2004potential}, and $D_0^{X}=4977.187(50)$~cm$^{-1}$~\cite{pashov2005potentials}. The present determination is certainly more accurate as it results from a direct difference of two frequencies, while the previous values were derived from the reconstruction of the PEC from spectroscopic data. 

\begin{figure}[!t]
	\centering
	\includegraphics[width=0.4\textwidth]{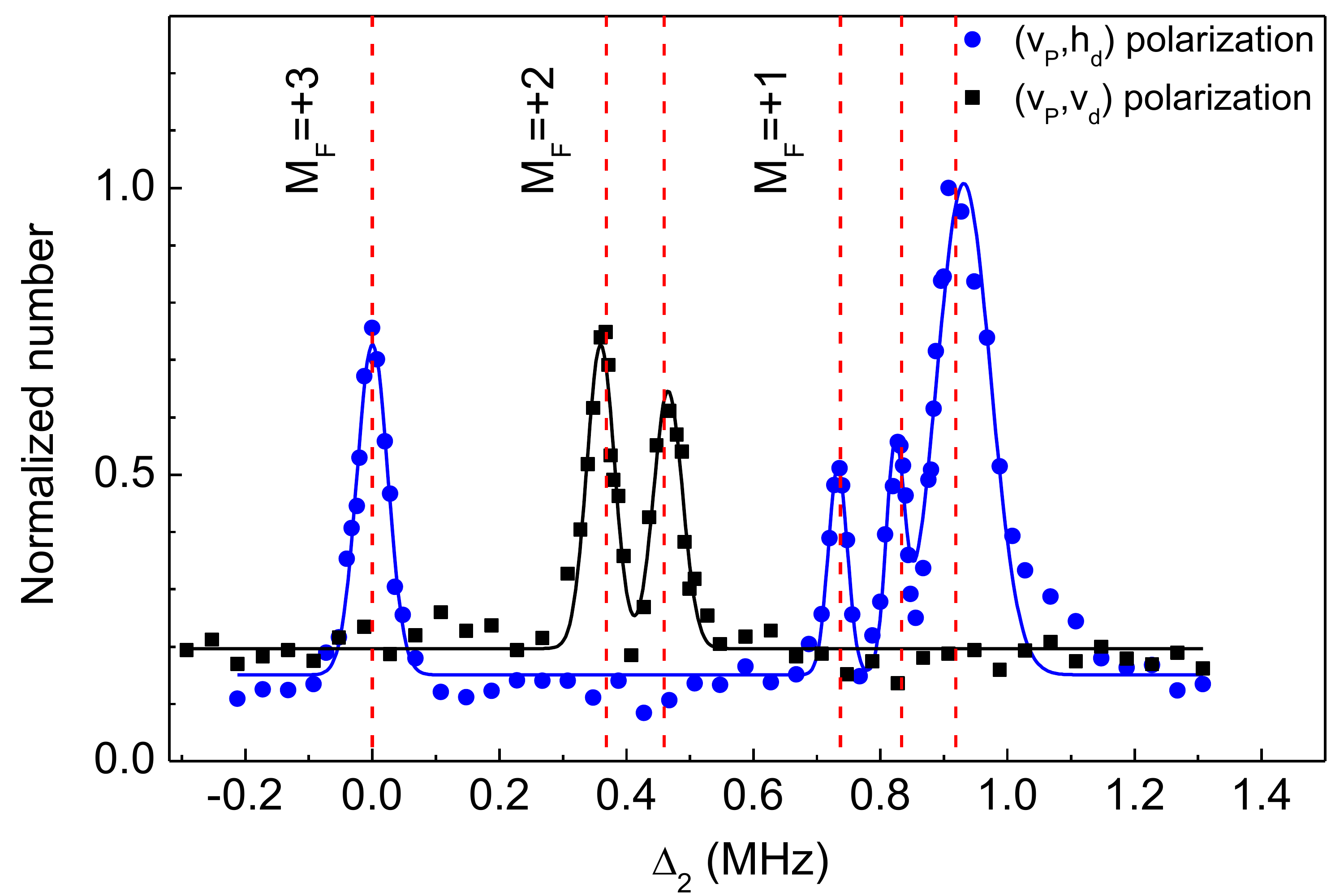}
	\caption{Two-photon spectrum of the $X^1\Sigma^+$ $\ket{v''=0,J''=0}$ level with resoled hyperfine structure. The Rabi frequency of the vertically polarized pump ($\rm{v}_{\rm p}$) laser $L_1$ is about $2\pi\times0.096$~MHz. Two $M''_F=+2$ levels are observed with vertically polarized dump ($\rm{v}_{\rm d}$, or $\pi$-polarized) laser $L_2$ with an intensity of 5.1~mW/cm$^2$; one $M_F=+3$ and three $M_F=+1$ hyperfine levels are observed with $L_2$ horizontally polarized dump ($\rm{h}_{\rm d}$) laser $L_2$ with an intensity of 51.0~mW/cm$^2$. The solid curves are Gaussian fits of the data for extracting the line centers, and the vertical dashed lines depict the theoretical predictions (Table~\ref{hyperfinecomponents}).}
	\label{fig11}
\end{figure}

\begin{table}[h]
	\centering
	\caption{Hyperfine structure of the $X^1\Sigma^+$ $\ket{v''=0,J''=0}$ level at 335.6~G: comparison of Theory (Th.) \cite{guo2016creation} and experiment (Exp.). The energy origin is set to the one of the absolute ground state with $M_F''=3$. Only six of the 16 sublevels are observed due to polarization selection rules. }
	\begin{tabular}{c|c|c|c|c}
		\hline\hline
		$M_F''$ & $M_{\rm Na}$ & $M_{\rm Rb}$ & Th.(MHz) & Exp.(MHz) \\
		\hline
		3 & 3/2 & 3/2 & 0.000 & 0.000(1) \\		
		2 & 1/2 & 3/2 & 0.368 & 0.359(1) \\		
		2 & 3/2 & 1/2 & 0.459 & 0.465(1) \\	
		1 & -1/2 & 3/2 & 0.737 & 0.733(2) \\	
		1 & 1/2 & 1/2 & 0.834 & 0.823(1) \\	
		1 & 3/2 & -1/2 & 0.918 & 0.931(2) \\	
		0 & -3/2 & 3/2 & 1.106 & - \\		
		0 & -1/2 & 1/2 & 1.208 & - \\		
		0 & 1/2 & -1/2 & 1.299 & - \\		
		0 & 3/2 & -3/2 & 1.377 & - \\		
		-1 & -3/2 & 1/2 & 1.583 & - \\	
		-1 & -1/2 & -1/2 & 1.679 & - \\	
		-1 & 1/2 & -3/2 & 1.764 & - \\	
		-2 & -3/2 & -1/2 & 2.060 & - \\	
		-2 & -1/2 & -3/2 & 2.151 & - \\	
		-3 & -3/2 & -3/2 & 2.537 & - \\		\hline\hline
	\end{tabular}	
	\label{hyperfinecomponents}
\end{table}  

The $\ket{v''=0,J''=0}$ level of the $X^1\Sigma^+$ state is actually split into $(2I_{\rm Na}+1)(2I_{\rm Rb}+1)=16$ hyperfine sublevels due to the atomic nuclear spins. At 335.6~G, the splitting between these structures is dominated by the nuclear Zeeman effect. We can thus label the hyperfine structures with $M_{\rm{Na}}$ and $M_{\rm{Rb}}$ for the projections of the nuclear spins along the magnetic field. Since the $X^1\Sigma^+$ state has no electronic spin and a vanishing projection of the electronic orbital angular momentum, the projection of the total molecular angular momentum $M_F''$ equals to $M_{\rm{Na}}+M_{\rm{Rb}}$. As listed in Table~\ref{hyperfinecomponents}, the $M_F''=3$ level is the absolute ground state for $^{23}$Na$^{87}$Rb. 

Despite the noticeable magnetic field of 335.6~G, the hyperfine structure is spread over 2.5~MHz only, while the spacings between adjacent structures are typically 10 times smaller. Our final goal being the transfer of the population into a single quantum level, namely the $M_F''=3$ one, the hyperfine structure must be resolved, requiring low Rabi frequencies (typically $2\pi\times 1$~MHz in our case) in order to keep a small two-photon linewidth. The resolved hyperfine structure in Fig.~\ref{fig11} is obtained with pulse length of 30~$\mu$s, and a $L_2$ laser power more than 100 times weaker than in Fig.~\ref{fig10}. Starting from the Feshbach state $\ket{1}$ with $M_F=+2$, with the polarization of $L_1$ set to vertical ($\pi$-light), only six hyperfine levels can be accessed with two different $L_2$ polarizations (Table~\ref{hyperfinecomponents}). The observed level spacings agree well with the calculated ones \cite{guo2016creation} using parameters from \textit{ab~initio} calculations for the hyperfine couplings~\cite{Aldegunde2017}.

\subsection{Calibration of the dump transition strength}

\begin{figure}[!t]
	\centering
	\includegraphics[width=0.4\textwidth]{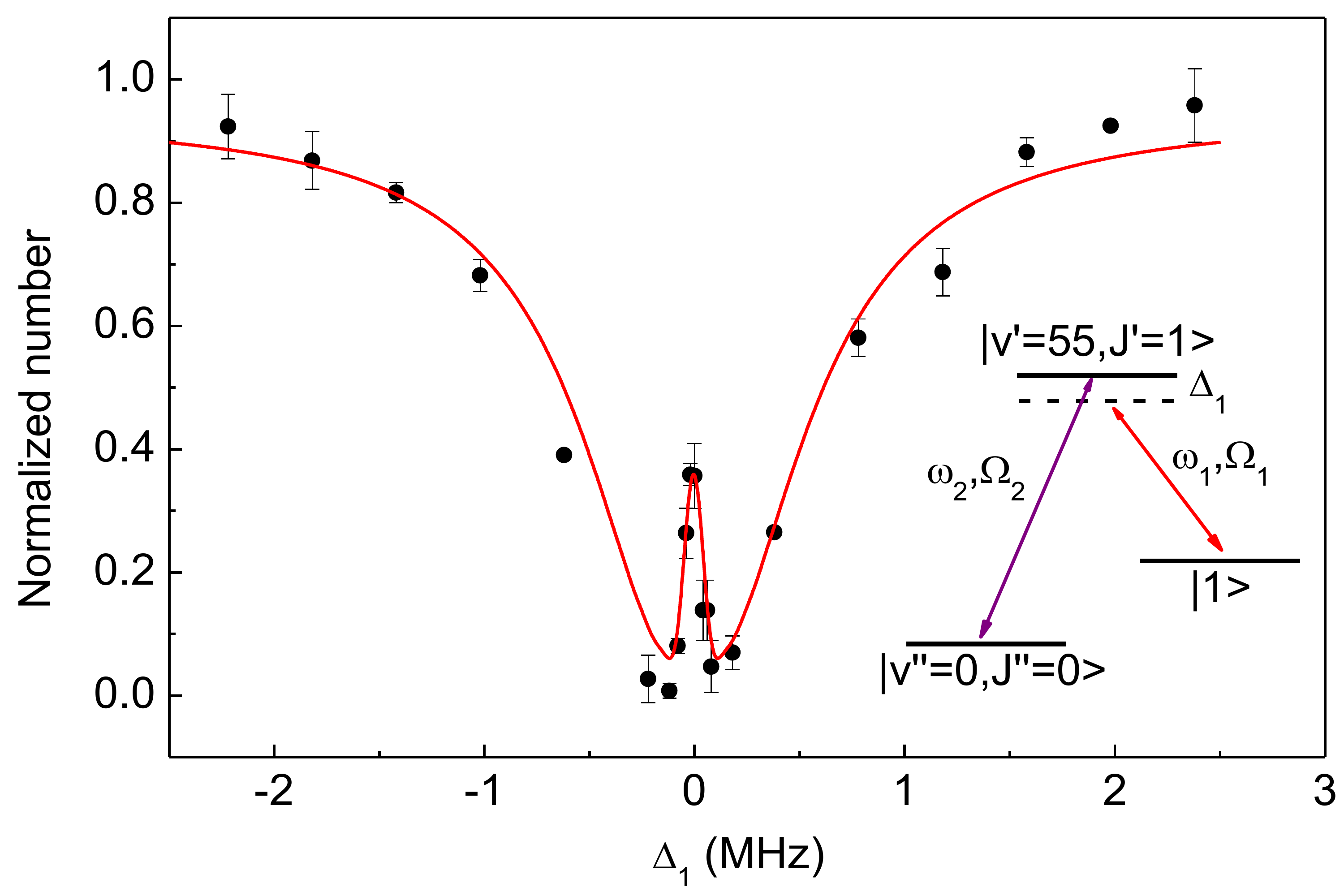}
	\caption{Two-photon dark-state spectroscopy for calibrating the dump transition strength. The $L_2$ frequency is kept on resonance and the $L_1$ frequency (and thus the detuning $\Delta_1$) is scanned. The red solid curve is the result of the fit of the signal using Eq.~(\ref{master}) with a pump Rabi frequency of $2\pi\times$0.169~MHz and the excited lifetime of $\tau$=238~ns. The dump Rabi frequency extracted from the fit is $2\pi\times$0.225~MHz, which corresponds to a normalized Rabi frequency of $\bar{\Omega}_2=2\pi\times$23.2 $\rm kHz\times\sqrt{I/(mW/cm^2)}$ and a TDM of 0.021~a.u.. }
	\label{fig12}
\end{figure}

To calibrate the coupling strength between the $\ket{2}$ and $\ket{3}$ levels, two-photon ``dark state'' spectroscopy is performed onto the hyperfine ground level of the $\ket{v''=0,J''=0}$ state~\cite{guo2016creation} with both $L_1$ and $L_2$ locked to the reference cavity. The $L_2$ laser is kept on resonance, while the frequency of $L_1$ is scanned with the AOM, as illustrated in the level scheme in Fig.~\ref{fig12}. When the two-photon resonance is satisfied, a small window of Feshbach molecule recovery shows up due to quantum interference even with rather low dump Rabi frequency. Figure~\ref{fig12} shows a typical ``dark state'' spectrum with the spacing between the two dips directly yielding the dump laser Rabi frequency. To avoid the influence of nearby hyperfine levels, the Rabi frequencies are relatively small which results in a non-perfect dark resonance due to the finite laser linewidths.

This three-level system can be modeled by the master equation \cite{whitley1976double} incorporating decoherences caused by both the spontaneous decay of the three levels and the finite linewidths of the two Raman lasers
\begin{equation}
\label{master}
	\dot{\rho}=-\frac{I}{\hbar}\space[H,\rho]-\frac{1}{2}\{\Gamma,\rho\} +L_\eta(\rho),
\end{equation}
with	$L_\eta(\rho)_{ij}=-\eta_{ij}\rho_{ij}$ ($i,j=1,2,3$). Here, $\rho$ is the density matrix, $H$ is the Hamiltonian of the system, $\Gamma$ is the matrix representing the decay rates of the three levels, and $\eta$ is the matrix representing the linewidths of the two lasers. Assuming that the two Raman lasers are independent, then we have $\eta_{13}=\eta_{12}+\eta_{23}$. Fitting the experimental ``dark state'' spectrum with Eq.~(\ref{master}) using our estimated laser linewidths of 5~kHz~\cite{di2010simple}, and the pump laser parameters from the aforementioned calibration, we obtain the normalized dump Rabi frequency $\bar{\Omega}_2=2\pi\times23.2$~kHz$\times\sqrt{\rm I/({mW/cm}^2)}$, and a TDM of 0.021~a.u.. Indeed, the dump transition is much stronger than the pump transition. This value is in remarkable agreement with the calculated one (0.034~a.u.) reported in the Appendix.

\section{Conclusion}

We have performed a detailed spectroscopic investigation of several selected vibrational levels of  $^{23}$Na$^{87}$Rb excited electronic states, exhibiting various characteristic patterns for their hyperfine structures. We developed a theoretical model for these hyperfine structures based on an effective Hamiltonian, which parameters were fitted to successfully reproduce the energy positions and intensities of the observed lines. We identified a specific level arising from the accidental coupling between two excited electronic states of $0^+$ and $0^-$ symmetry, ideal for transferring $^{23}$Na$^{87}$Rb Feshbach molecules to the absolute ground state with STIRAP. After careful calibrations of both the pump and dump transition strengths, a high efficiency STIRAP transfer is demonstrated~\cite{guo2016creation}, which allowed us to produce a high density sample of polar $^{23}$Na$^{87}$Rb molecules in their absolute ground state. As a side result, we also obtained a significant discrepancy between the measured and the computed lifetime of the selected levels, assigned to its predissociation through the repulsive branch of the lowest triplet electronic state, thus competing with radiative decay. This coupling, often referred to as heterogeneous perturbation~\cite{lefebvre2004spectra} and invoked for instance in Refs.~\cite{schmidt-mink1985b,kowalczyk2015direct}, is not taken in account in our model.

Given the density of rovibrational levels in heavy alkali dimers, it is likely that the accidental coupling of the $0^+$ and $0^-$ electronic states could be detected in other species as well, thus offering a robust STIRAP scheme. However the prediction of the energy location of such coupled levels is hopeless without a careful investigation of the related $A-b_{0^+}$ system. Note however that their spectroscopy has been investigated for several heteronuclear alkali pairs, so that a careful reexamination of the data could reveal such an accidental perturbation.

\begin{acknowledgments}
	This work is supported by the COPOMOL project which is jointly funded by Hong Kong RGC (grant no. A-CUHK403/13) and France ANR (grant no.ANR-13-IS04-0004-01). The Hong-Kong team is also supported by the RGC General Research Fund (grant no.CUHK404712) and the National Basic Research Program of China (grant No. 2014CB921403).
\end{acknowledgments}

\begin{widetext}
\appendix*
\section{}

\subsection{Matrix elements of the effective Hamiltonian}

Explicit expressions of the matrix elements of the effective Hamiltonian terms are given in the uncoupled basis $\ket{J\Omega M_JM_{\rm{Na}}M_{\rm{Rb}}}$. We considered the magnetic dipole interaction (rank 1 hyperfine operators), the electric quadrupole interaction (rank 2 hyperfine operator), the Zeeman effect and the rotational operator. 

\subsubsection{Hyperfine operators}

Following \cite{broyer1978}, the hyperfine operators can be written as a linear combination of $Q^k_qV^k_{-q}$ with $Q^k_q$ (resp. $V^k_{-q}$) being a spherical tensor of rank k acting on the nuclear coordinate (resp. electronic coord.). Working on a decoupled basis set, the matrix elements are directly
\begin{equation}
 \braket{i'|Q^k_qV^k_{-q}|i}=\braket{M_{\rm{Na}}'M_{\rm{Rb}}'|Q^k_q|M_{\rm{Na}}M_{\rm{Rb}}}\braket{J'\Omega' M'_J|V^k_{-q}|J\Omega M_J}.
\end{equation}
The following equations are derived for Na and are obviously identical for Rb. Using the Wigner-Eckart theorem on the nuclear part we get for an operator acting on the nucleus Na
\begin{equation}
 \braket{M_{\rm{Na}}'|Q^k_q|M_{\rm{Na}}} = (-1)^{I_{\rm{Na}}-M_{\rm{Na}}} \left( \begin{array}{ccc} I_{\rm{Na}} & k & I_{\rm{Na}} \\ -M_{\rm{Na}}' & q & M_{\rm{Na}}\end{array} \right) \braket{I||Q^k_q||I},
\end{equation}
where (...) is a 3j-coefficient, which is non-zero only for $q=M_{\rm{Na}}'-M_{\rm{Na}}$. Note that the latter $q$ integer is not identical to the $q$ character present in the standard notation \cite{broyer1978} $eqQ$ for the nuclear quadrupole interaction.

The reduced matrix element is given by
\begin{equation}
\braket{I||Q^k_q||I} = \braket{I M_I=I|Q^k_q||I M_I=I} \left( \begin{array}{ccc} I_{\rm{Na}} & k & I_{\rm{Na}} \\ -I_{\rm{Na}} & 0 & I_{\rm{Na}}\end{array} \right)^{-1}.
\end{equation}
For rank 1 operators, $\braket{II|Q^k_q||II}$ is usually defined as the magnetic moment $g_I\mu_NI$ with $g_I$ the nuclear $g$ factor and $\mu_N$ the nuclear magneton. For the rank 2 operator the matrix element is the nuclear quadrupole moment $\frac{1}{2}eQ$. A similar treatment on the electronic part of the operators gives

\begin{equation}
  \braket{J'\Omega' M'_J|V^k_{-q}|J\Omega M_J} \propto (-1)^{-\Omega'-M_J'}\sqrt{(2J+1)(2J'+1)} \left( \begin{array}{ccc} J' & k & J \\
-M_J' & -q & M_J\end{array} \right) \left( \begin{array}{ccc} J' & k & J \\
-\Omega' & \Delta\Omega & \Omega  \end{array} \right).
\end{equation}

We finally obtain for the magnetic dipole and electric quadrupole operators acting on Na
\begin{align}
\braket{i'|H_{\rm{mag}}({\rm{Na}})|i}&= w_{\rm{Na}} I_{\rm{Na}} \braket{i'|\textbf{H}_{hfs}^{(1)}|i};\\
\braket{i'|H_{\rm{eqQ}}({\rm{Na}})|i}&= \frac{1}{2}eqQ \braket{i'|\textbf{H}_{\rm{hfs}}^{(2)}|i} ,
\end{align}
with $w_{\rm{Na}}$ a linear combination of the three magnetic dipole coupling constants ($G_{LI}$,$D_{SI}$,$K_{FI}$) and
\begin{align}
\braket{J'\Omega'M_J'M_{\rm{Na}}'M_{\rm{Rb}}'|\textbf{H}_{\rm{hfs}}^{(k)}({\rm{Na}})|J\Omega M_JM_{\rm{Na}}M_{\rm{Rb}}}&= (-1)^{I_{\rm{Na}}-\Omega-M_J-M_{\rm{Na}}} \delta_{M_{\rm{Rb}}' M_{\rm{Rb}}} \sqrt{(2J+1)(2J'+1)} \notag\\& \left( \begin{array}{ccc} J' & k & J \\
-M_J' & -q & M_J\end{array} \right) \left( \begin{array}{ccc} J' & k & J \\
-\Omega' & \Delta\Omega & \Omega  \end{array} \right)\notag\\& \left( \begin{array}{ccc} I_{\rm{Na}} & k & I_{\rm{Na}} \\
-M_{\rm{Na}}' & q & M_{\rm{Na}}\end{array} \right) \left( \begin{array}{ccc} I_{\rm{Na}} & k & I_{\rm{Na}} \\
-I_{\rm{Na}} & 0 & I_{\rm{Na}}\end{array} \right)^{-1}.
\end{align}
The same derivation holds for the operators acting on the Rb nucleus.

\subsubsection{Zeeman operator}

Matrix element of the Zeeman effect with the laboratory z-axis defined as the magnetic field axis
\begin{align}
\braket{J'\Omega M_J M_{\rm{Na}}'M_{\rm{Rb}}'|H_{\textrm{Zeeman}}|J\Omega M_JM_{\rm{Na}}M_{\rm{Rb}}}&= (-1)^{-\Omega'-M_J'} \frac{\mu_B}{\hbar} (g_L\Lambda +g_S\Sigma) B \sqrt{(2J+1)(2J'+1)} \nonumber\\& \left( \begin{array}{ccc} J' & 1 & J \\
-M_J & 0 & M_J\end{array} \right) \left( \begin{array}{ccc} J' & 1 & J \\
-\Omega & 0 & \Omega  \end{array} \right) .
\end{align}

\subsubsection{Rotational operator}

In the Hund's case (c) basis we have
\begin{equation}
\braket{J\Omega M_JM_{\rm{Na}}M_{\rm{Rb}}|H_{\textrm{Rot}}|J\Omega M_JM_{\rm{Na}}M_{\rm{Rb}}}= B_v (J(J+1) - 2\Omega^2).
\end{equation}
Note that off-diagonal rotational couplings were not considered in this work.

\subsection{Transition dipole moment}
\label{tdm}

We recall here the expression of the transition dipole moment between a Feshbach level and the excited level. One difficulty for computation is that we need to express the wavefunction of both level in the same basis. After using the effective Hamiltonian the wavefunction of the excited level is obtained as a linear combination of Hund's case (c) basis vectors. As the mixing of $0^+$ due to the SO coupling is known, we can further decomposed the wavefunction into a spin-uncoupled Hund's case (a) basis :
\begin{equation}
 \ket{\Psi_e} = \sum_{\alpha} c_{\alpha} \ket{J_{\alpha} \Lambda_{\alpha} \Sigma_{\alpha} M_J^{\alpha} ; I_{\rm{Na}} M_{\rm{Na}}^{\alpha} ; I_{\rm{Rb}} M_{\rm{Rb}}^{\alpha}}.
\end{equation}
The wavefunction of a Feshbach molecule is usually expressed in Hund's case (e), the angular momenta of the electrons being only weakly coupled to the internuclear axis. The wavefunction can be expressed in a molecular basis \cite{gaoren-wang-pc}
\begin{equation}
\ket{\Psi_f}=\sum_{SM_SIM_I}c_{\beta}\ket{lm_l;SM_SIM_I}.
\end{equation}
Assuming we have for the ground state (s+s) $l \equiv N$, this basis is equivalent to a spin-coupled Hund's case (b) basis. Using angular algebra we expressed it in the spin-uncoupled basis
\begin{equation}
 \ket{lm_l;SM_SIM_I}=\sum_{M_{\rm{Na}}M_{\rm{Rb}}} (-1)^{I_{\rm{Na}}-I_{\rm{Rb}}+M_I} \left( \begin{array}{ccc} I_{\rm{Na}} & I_{\rm{Rb}} & I \\ M_{\rm{Na}} & M_{\rm{Rb}} & -M_I  \end{array} \right) \ket{N S m_l M_S ; I_{\rm{Na}} M_{\rm{Na}} ; I_{\rm{Rb}} M_{\rm{Rb}}}.
\end{equation}
The wavefunction is then expressed in the Hund's case (a) basis
\begin{align}
 \ket{lm_l;SM_SIM_I}&=\sum_{\Sigma,J,M_{\rm{Na}}M_{\rm{Rb}}} (-1)^{J-l-M_J-\Lambda+I_{\rm{Na}}-I_{\rm{Rb}}+M_I} \sqrt{(2J+1)(2l+1)}
 \left( \begin{array}{ccc} J & S & l \\ -\Omega & \Sigma & \Lambda  \end{array} \right) 
 \nonumber\\&
 \left( \begin{array}{ccc} S & l & J \\ M_S & m_l & -M_J  \end{array} \right) \left( \begin{array}{ccc} I_{\rm{Na}} & I_{\rm{Rb}} & I \\ M_{\rm{Na}} & M_{\rm{Rb}} & -M_I  \end{array} \right)
 \ket{J \Lambda\Sigma M_J ; I_{\rm{Na}} M_{\rm{Na}} ; I_{\rm{Rb}} M_{\rm{Rb}} } .
\end{align}
Finally, taking the matrix element of the dipole operator in the Hund's case (a) basis, we have
\begin{equation}
  \begin{split}
  \braket{\Psi_e | \mu | \Psi_f} &= \left| \sum_{pq,\alpha,SM_SIM_I,\Sigma,J,M_{\rm{Na}}M_{\rm{Rb}}} \delta_{\Sigma\Sigma_{\alpha}} \delta_{M_{\rm{Na}}M_{\rm{Na}}^{\alpha}}\delta_{M_{\rm{Rb}}M_{\rm{Rb}}^{\alpha}} c_{\alpha}c_{\beta} d_q e_p (-1)^{-\Omega_{\alpha}-M_J^{\alpha} +J-l-M_J-\Lambda+I_{\rm{Na}}-I_{\rm{Rb}}+M_I} \right.
  \nonumber\\ 
  &(2J+1) \sqrt{(2J_{\alpha}+1)(2l+1)} 
  \left( \begin{array}{ccc} J & S & l \\ -\Omega & \Sigma & \lambda  \end{array} \right) \left( \begin{array}{ccc} S & l & J \\ M_S & m_l & -M_J  \end{array} \right) 
  \nonumber\\& 
  \left.
  \left( \begin{array}{ccc} I_{\rm{Na}} & I_{\rm{Rb}} & I \\ M_{\rm{Na}} & M_{\rm{Rb}} & -M_I  \end{array} \right) 
  \left( \begin{array}{ccc} J_{\alpha} & 1 & J \\ -M_J^{\alpha} & p & M_J  \end{array} \right)  \left( \begin{array}{ccc} J_{\alpha} & 1 & J \\ -\Lambda_{\alpha} & q & \Lambda  \end{array} \right) \right|,
   \end{split}
\end{equation}
with $e_p$ linked to the polarization of the laser ($e_0=e_z$ and $e_{\pm1}=\mp \frac{e_x \pm i e_y}{\sqrt{2}}$) and $d_q$ the electronic and vibrational transition dipole moment ($d_0=d_z$ and $d_{\pm1}=\mp \frac{d_x \pm i d_y}{\sqrt{2}}$). The electric dipole operator does not act on the spin, meaning only components with the same spin S and spin projection $\Sigma$, $M_{\rm{Na}}$ and $M_{\rm{Rb}}$ for both levels are coupled.

We note that the wavefunction of the absolute ground level ($X^1\Sigma^+$ ;$v=0$) is also expressed in the Hund's case (b) basis, meaning a similar expression can be used to compute the dump transition dipole moment.

\subsection{The hyperfine components of the $\ket{v'=55, J'=1}$ level}
\label{v55}

In Table~\ref{tab55}, we reported the main features of all the hyperfine components of $\ket{v'=55, J'=1}$ level that can be reached from the $\ket{1}$ Feshbach level, and coupled to the $\ket{3}$ absolute ground state level (Fig.~\ref{fig8}). The listed energies are obtained from the fitting procedure with respect to the energy position of the barycenter of the manifold. It is clear that the central lines located between -54~MHz and -8.8~MHz correspond to almost pure $0^+$ levels grouped into a narrow hyperfine structure. In contrast, the lateral lines exhibit strongly mixed $0^+$ and $0^-$ characters, as confirmed by the two main basis vectors contributing to their wave function. 

As an illustration, it is worthwhile to display the full composition of the selected level $\ket{2}$ (highlighted with horizontal lines in the Table):  
$\ket{0^-, 0, 0, 3/2, 1/2}$ (27.63\%), 
$\ket{0^+, 1, 0, 3/2, 1/2}$ (22.58\%), 
$\ket{0^+, 1, 0, 1/2, 3/2}$ (19.46\%), 
$\ket{0^-, 0, 0, 1/2, 3/2}$ (15.73\%), 
$\ket{0^+, 1, 1, 3/2,-1/2}$ (9.63\%),
$\ket{0^+, 1,-1, 3/2, 3/2}$ (3.05\%), 
$\ket{0^+, 1, 1,-1/2, 3/2}$ (1.29\%),
$\ket{0^+, 1, 1, 1/2, 1/2}$ (0.47\%).
The remaining 0.16\% concern $0^-$ vectors with $J'=2$.

The TDMs for the pump transitions have been measured for most of the lines of the lateral bands in (Fig.~\ref{fig8}), and their values are consistent with those yielded by the theoretical model. As reported in the text, we see in particular that the TDM of the pump transition is in general much smaller than the one of the dump transition.

\subsection{The radiative lifetimes of the $\ket{v'=55, J'=1}$ sublevels}
\label{lifetimes}

Another noticeable feature which has been briefly addressed in the text concerns the computed radiative lifetimes which are significantly larger than the one measured for the selected $\ket{2}$ level. They are calculated as follows in our model. The initial coupled channel calculations gave the magnitude of the singlet and triplet components of the $(A-b_{0^+}) v'=55$ vibrational level (Table \ref{components}). The level can thus decay via spontaneous emission either towards a vibrational level of the $X^1\Sigma^+$ singlet ground state, or to the
$a^3\Sigma^+$ metastable triplet state. We have computed the related Einstein coefficients $A_X$ and $A_a$ for all the dipole-allowed transitions, taking into account both bound-bound and bound-free contributions. Summing over all possible final levels, we obtained $A_X=2.97 \times 10^6$~s$^{-1}$ (dominated by bound-bound transitions) and $A_a=2.05\times 10^5$~s$^{-1}$ (mainly bound-free transitions). The total Einstein coefficient $A_{0^+}$ for this level is thus $A_{0^+}= 3.18\times 10^6$~s$^{-1}$, leading to a lifetime of ~315 ns. Assuming that the $\Omega=0^-$ level can be represented by a pure $^3\Pi$ wavefunction we obtain with a similar procedure $A_{0^-}=2.06\times 10^5$~s$^{-1}$. Due to the hyperfine interaction, each sublevel has a mixed $0^+,0^-$ character (partially reported in Table~\ref{tab55}). As the spontaneous emission does not depend on the angular and hyperfine parts of the wavefunction, we can, in a first approximation, obtain the Einstein coefficient for each sublevel by taking the weighted sum of the corresponding $0^+$ and $0^-$ values.

As quoted previously, the sublevel selected for STIRAP has a measured lifetime of 238~ns, shorter than the theoretical value of 535~ns. Following the same procedure for the (almost pure triplet) $\Omega=1,v'=60$ level, we found an even longer lifetime (4372~ns) compared to the experimental one (404~ns). This discrepancy could be explained by the fast predissociation of the $b^3\Pi$ levels toward the $a^3\Sigma^+$ radial continuum due to rotational coupling at short internuclear distances which is not taken into account in our theoretical model. In this regard we note that our results with $\Omega=1$ and $\Omega=0$ are consistent with each other. Indeed, if we replace the computed Einstein coefficient for all triplet-triplet transition by a fixed predissociation rate of $2.47\times 10^6$~s$^{-1}$ then the computed lifetimes are 242~ns and 405~ns for the $(A-b_{0^+}) v'=55$ and $\Omega=1,v'=60$ levels, close to the measured values.

\begin{table}[!h]
   \begin{center}
	\caption{The main features of all the hyperfine components of $\ket{v'=55, J'=1}$ level resulting from the fitting procedure: $M'_F$, energies, percentage of $0^+$ component, radiative lifetime, pump TDM compared to the measured values, dump TDM, and the two main basis vectors of the related wave functions with their weights rounded off to the closest integer value.}
\begin{tabular}{|c|c|c|c|c|c|c|c|c|c|c|} 
\hline
$M'_F$ & Energy & $0^+$ & Lifetime & Computed & Experimental & Computed & \multicolumn{2}{c|}{Main character} & \multicolumn{2}{c|}{2$^{nd}$ main character} \\
\cline{8-11}
 & (MHz) & char.  &(ns)& Pump tdm & Pump tdm & Dump tdm &  & Weight &  & Weight\\
 &  &(\%) & &($10^{-4}$~a.u.) & ($10^{-4}$~a.u.) & (a.u.)  &$\ket{\Omega, J, M_J, M_{{Na}}, M_{{Rb}}}$ & (\%)&$\ket{\Omega, J, M_J, M_{\rm{Na}}, M_{\rm{Rb}}}$ &(\%)\\
\hline
 3 &-956.49 & 55.8 & 536.7 & 2.4 &0.745 	&0.13606 	& $\ket{0^+, 1, 0, \frac{3}{2}, \frac{3}{2}}$ & 47	& $\ket{0^-, 0, 0,\frac{3}{2}, \frac{3}{2}}$ & 44 \\
 2 &-897.21 & 55.9 & 535.3 & 3.9 &5.5  	&0.05693 	& $\ket{0^-, 0, 0, \frac{1}{2}, \frac{3}{2}}$ & 28 	& $\ket{0^+, 1, 0,\frac{1}{2}, \frac{3}{2}}$ & 24 \\
 1 &-833.35 & 56.2 & 533.4 & 2.7 &2.58 	&0. 		& $\ket{0^-, 0, 0, \frac{1}{2}, \frac{1}{2}}$ & 24 	& $\ket{0^-, 0, 0, -\frac{1}{2}, \frac{3}{2}}$ & 15 \\ \hline
 2 &-733.33 & 56.5 & 530.9 & 4.7 &6.97 	&0.03466 	& $\ket{0^-, 0, 0, \frac{3}{2}, \frac{1}{2}}$ & 28 	& $\ket{0^+, 1, 0,\frac{3}{2}, \frac{1}{2}}$ & 23 \\ \hline
 1 &-666.87 & 56.8 & 528.0 & 3.8 & 3.46	&0. 		& $\ket{0^-, 0, 0, -\frac{1}{2}, \frac{3}{2}}$ & 24 	& $\ket{0^+, 1, 0, -\frac{1}{2}, \frac{3}{2}}$ &24 \\
 1 &-508.58 & 58.2 & 516.8 & 4.1 & 2.98	&0. 		& $\ket{0^-, 0, 0, \frac{3}{2}, -\frac{1}{2}}$ & 24 	& $\ket{0^-, 0, 0, \frac{1}{2}, \frac{1}{2}}$ &14 \\
 3 & -54.00 & 99.5 & 316.2 & 2.4 & -		&0.06832 	& $\ket{0^+, 1, 1, \frac{1}{2}, \frac{3}{2}}$ & 64 	& $\ket{0^+, 1, 1, \frac{3}{2},\frac{1}{2}}$ & 24 \\
 2 & -49.63 & 99.6 & 316.0 & 2.0 & -		&0.00711 	& $\ket{0^+, 1, 1, -\frac{1}{2}, \frac{3}{2}}$ & 36 	& $\ket{0^+, 1, 1, \frac{1}{2},\frac{1}{2}}$ & 32 \\
 1 & -45.10 & 99.6 & 315.9 & 1.5 & -		&0.		& $\ket{0^+, 1, 1, -\frac{1}{2}, \frac{1}{2}}$ & 33 	& $\ket{0^+, 1, 0, \frac{1}{2}, \frac{1}{2}}$ & 23\\
 2 & -43.12 & 99.7 & 315.8 & 3.8 & -		&0.1523	& $\ket{0^+, 1, -1, \frac{3}{2}, \frac{3}{2}}$ & 59 	& $\ket{0^+, 1, 1,\frac{1}{2}, \frac{1}{2}}$ & 20 \\
 3 & -40.66 & 99.7 & 315.9 & 3.6 & -		&0.04059 	& $\ket{0^+, 1, 1, \frac{3}{2}, \frac{1}{2}}$ & 63 	& $\ket{0^+, 1, 1, \frac{1}{2},\frac{3}{2}}$ & 33 \\
 1 & -38.72 & 99.7 & 315.6 & 1.6 & -		&0. 		& $\ket{0^+, 1, -1, \frac{1}{2}, \frac{3}{2}}$ & 46 	& $\ket{0^+, 1, 1, \frac{1}{2}, -\frac{1}{2}}$ & 25\\
 2 & -33.87 & 99.7 & 315.8 & 3.5 & -		&0.06965 	& $\ket{0^+, 1, 0, \frac{3}{2}, \frac{1}{2}}$ & 33 	& $\ket{0^+, 1, 1, -\frac{1}{2},\frac{3}{2}}$ & 28 \\
 1 & -28.81 & 99.7 & 315.6 & 2.8 & -		&0. 		& $\ket{0^+, 1, -1, \frac{1}{2}, \frac{3}{2}}$ & 19 	& $\ket{0^+, 1, 1, -\frac{1}{2}, \frac{1}{2}}$ & 17\\
 1 & -25.00 & 99.8 & 315.4 & 0.5 & -		&0. 		& $\ket{0^+, 1, 0, \frac{3}{2}, -\frac{1}{2}}$ & 38 	& $\ket{0^+, 1, -1, \frac{3}{2}, \frac{1}{2}}$ & 34\\
 2 & -23.53 & 99.8 & 315.5 & 4.8 & -		&0.05756 	& $\ket{0^+, 1, 1, \frac{3}{2}, -\frac{1}{2}}$ & 39 	& $\ket{0^+, 1, 1,\frac{1}{2}, \frac{1}{2}}$ & 29 \\
 1 & -17.62 & 99.8 & 315.4 & 2.5 & -		&0. 		& $\ket{0^+, 1, 1, -\frac{3}{2}, \frac{3}{2}}$ & 32 	& $\ket{0^+, 1, 0, \frac{3}{2}, -\frac{1}{2}}$ & 28\\
 1 &  -8.81 & 99.9 & 315.2 & 1.7 & -		&0. 		& $\ket{0^+, 1, 1, \frac{3}{2}, -\frac{3}{2}}$ & 38 	& $\ket{0^+, 1, 1, \frac{1}{2}, -\frac{1}{2}}$ & 17\\
 1 & 674.79 & 41.6 & 693.0 & 4.2 & 4.18	&0. 		& $\ket{0^-, 0, 0, \frac{3}{2}, -\frac{1}{2}}$ & 35 	& $\ket{0^-, 0, 0, \frac{1}{2}, \frac{1}{2}}$ & 18\\
 1 & 821.72 & 42.9 & 675.5 & 3.7 & 3.56	&0. 		& $\ket{0^-, 0, 0, -\frac{1}{2}, \frac{3}{2}}$ & 31 	& $\ket{0^-, 0, 0, \frac{3}{2}, -\frac{1}{2}}$ &18 \\
 2 & 880.97 & 43.2 & 671.5 & 5.2 & 4.18	&0.03237 	& $\ket{0^-, 0, 0, \frac{3}{2}, \frac{1}{2}}$ & 38 	& $\ket{0^-, 0, 0,\frac{1}{2}, \frac{3}{2}}$ & 19 \\
 1 & 966.54 & 43.4 & 668.0 & 2.5 & 2.02	&0. 		& $\ket{0^-, 0, 0, \frac{1}{2}, \frac{1}{2}}$ & 31 	& $\ket{0^-, 0, 0, -\frac{1}{2}, \frac{3}{2}}$ & 20\\
 2 & 1022.4 & 43.6 & 666.0 & 4.1 & 3.5 	&0.04799	& $\ket{0^-, 0, 0, \frac{1}{2}, \frac{3}{2}}$ & 37 	& $\ket{0^+, 1, 0, \frac{1}{2},\frac{3}{2}}$ & 20 \\
 3 & 1073.4 & 43.7 & 664.6 & 2.2 & 1.93	&0.11968 	& $\ket{0^-, 0, 0, \frac{3}{2}, \frac{3}{2}}$ & 56 	& $\ket{0^+, 1, 0,\frac{3}{2}, \frac{3}{2}}$ & 36 \\ 
\hline	 
 
\end{tabular}
\label{tab55}	
   \end{center}
\end{table}

\end{widetext}

\end{document}